# A review of simulation, measurement techniques, and development in chip thermal design*

ZHOU Junnian, ZHOU Feng, YUE Shengying*

State Key Laboratory for Strength and Vibration of Mechanical Structures, Xi'an Jiaotong University, Xi'an 710049, China

**Abstract** As integrated circuits advance toward higher power densities, three-dimensional integration, and heterogeneous packaging, chip thermal management has become a key bottleneck limiting device performance, reliability, and lifetime. This article systematically reviews numerical simulation methods and experimental measurement techniques for chip thermal design, with particular emphasis on the technical challenges associated with multiscale and multiphysics coupling, thermal boundary resistance measurement, and high-heat-flux cooling. We first introduce macro- and device-scale thermal simulation methods, including equivalent thermal-circuit models, the finite element method, and computational fluid dynamics, and discuss the application of phonon transport theory and molecular dynamics at microscopic scales. We then examine the advantages and limitations of infrared thermography, thermoreflectance, Raman thermometry, and embedded sensors. Current limitations include the enormous computational cost, inaccurate multiscale coupling, expensive experimental facilities, and the physical limits of conventional cooling technologies. Finally, we discuss emerging directions, including AI-accelerated thermal simulation, embedded microchannel liquid cooling, two-phase cooling, advanced high-thermal-conductivity materials, and multiphysics co-design, with the aim of advancing chip thermal management toward greater efficiency and intelligence.

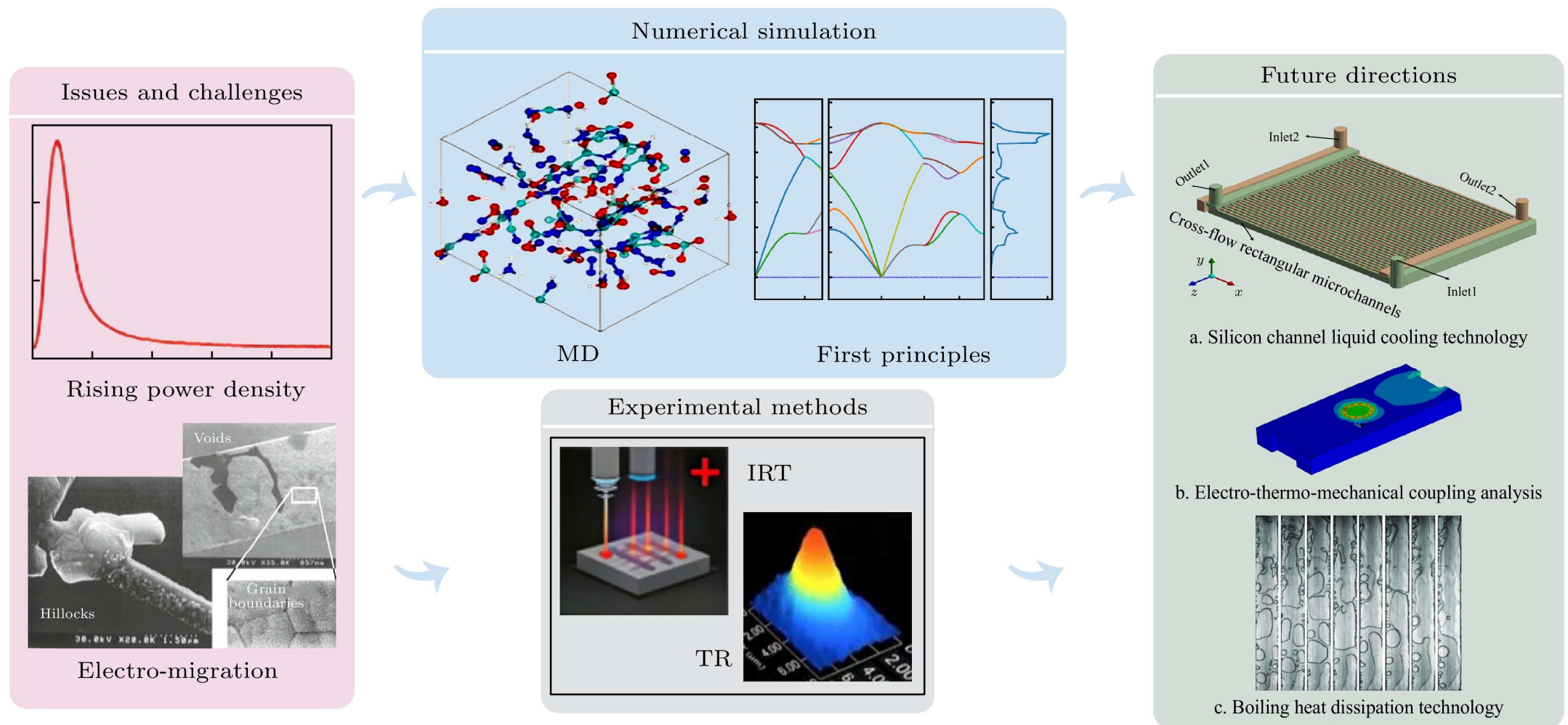

# 1 Introduction

As Moore's law slows and semiconductor technology increasingly shifts toward three-dimensional integration, system-on-chip (SoC) designs, and heterogeneous packaging, chip power density continues to rise rapidly. The transition from conventional two-dimensional devices to 2D+, 2.5D/3D packaging, chiplet architectures, and heterogeneous integration has made package structures more complex and interconnects more compact. Local hotspot heat fluxes have exceeded 1000 W/cm$^2$, making thermal management a primary bottleneck to further improvements in performance, reliability, and lifetime. In multilayer-stacked 3D integrated circuits, for example, constrained heat-transfer channels and increased interlayer thermal resistance make thermal design substantially more difficult[1,2].

From the perspective of the underlying physics, high internal temperatures intensify failure mechanisms such as thermal stress, thermomigration, electromigration, and negative-bias temperature instability (NBTI), thereby posing severe challenges to device reliability. Differences in the coefficients of thermal expansion of the constituent materials can generate shear deformation, cracking, and interfacial mismatch. Thermomigration can cause metal interconnects or through-silicon vias (TSVs) to migrate or fracture, while electromigration and NBTI accelerate at elevated temperatures and, in extreme cases, lead to transistor or interconnect failure. Thermal design therefore concerns not only heat-removal capacity but also electrical performance, electro-thermo-mechanical coupling, structural integrity, and system lifetime.
Thermal design based on experience or coarse empirical models can no longer meet the demands of today's high-density, high-power chips. Accurate numerical simulation and high-resolution experimental validation have become essential. Simulation tools predict temperature fields, heat-flow paths, and structural stresses, whereas experiments validate models, measure thermophysical properties, and calibrate boundary conditions and power maps, thereby improving design accuracy and reliability. Only by combining the two can the challenges posed by evolving package structures, complex heat-dissipation paths, and strong thermo-mechanical coupling be addressed.
Accordingly, this article systematically reviews the principal numerical simulation methods and experimental measurement techniques used in chip thermal design, analyzes the technical bottlenecks associated with high-density and 3D integration, and discusses future research directions. Section 2 introduces numerical simulation methods; Section 3 reviews experimental measurement and simulation-validation techniques; Section 4 analyzes current technical bottlenecks; Section 5 discusses potential research directions; and Section 6 summarizes the main conclusions and the central role of thermal design in future chips. The review is intended to provide a systematic, forward-looking, and practical reference for researchers and engineers working on chip thermal management, package cooling, numerical simulation, and thermal measurement. It spans macroscopic simulation, microscopic transport mechanisms, experimental validation, and system-level cooling design, with the goal of supporting advances in thermal management for next-generation high-power-density chips.

# 2 Numerical Simulation

## 2.1 Chip- and Package-Level Thermal Simulation Methods

At chip and package scales, heat conduction is still governed primarily by Fourier's law, which can be written as

$$\rho C_{\mathrm{p}} \frac{\partial T}{\partial t} = \nabla \cdot (k \nabla T) + Q(r,t), \tag{1}$$

Here, Q is obtained from the power map and is a key input to chip-level simulations. On the basis of this equation, equivalent thermal-circuit models have become the principal means of rapidly evaluating temperature fields at the architecture level. These models map chip material layers onto RC networks of thermal resistances and capacitances, so that the solution resembles a transient circuit analysis. HotSpot is a representative example and can predict steady-state and transient temperatures of processors or accelerators within a short time; it is therefore widely used in early microarchitecture design, floorplanning, and the optimization of dynamic thermal-management strategies[3]. As 3D ICs and chiplet systems have developed, however, thermal paths have become much more complex. Simple RC networks cannot adequately resolve variations in through-silicon-via (TSV) density, thermal boundary resistance (TBR), or stacked-package geometry, and engineering analyses therefore often require calibration against higher-fidelity simulations.

The finite element method (FEM) and finite volume method (FVM) are the principal high-fidelity tools for package-level thermal simulation. By discretizing complex package geometries, they can resolve multiscale structures, interface conditions, material heterogeneity, and external cooling environments. For example, FEM heat-conduction models of high-bandwidth-memory (HBM) stacks can predict interlayer heat-flow paths and high-heat-flux regions around TSVs when material properties are treated as uniform, interfacial thermal resistances are known, and the chip layers and TSVs dominate heat spreading. Under identical boundary conditions and power inputs, comparison with infrared-thermography measurements has shown a maximum relative deviation below 5%, supporting the applicability and accuracy of FEM for HBM package analysis[4]. When a model expands to hundreds of millions of elements, however, a full transient solution may require days or even weeks, which is a typical limitation of FEM/FVM in 3D-IC thermal design.

For systems employing air cooling, cold plate liquid cooling, or immersion cooling, thermal issues involve not only solid heat conduction but also convective heat transfer within fluids. In such cases, it is necessary to solve fluid dynamics models coupling the Navier-Stokes equations with the energy equation:

$$\begin{aligned} \rho\left(\frac{\partial \boldsymbol{u}}{\partial t} + \boldsymbol{u} \cdot \nabla \boldsymbol{u}\right) &= -\nabla p + \mu \nabla^2 \boldsymbol{u}, \\ \rho C_{\mathrm{p}}(\boldsymbol{u} \cdot \nabla T) &= \nabla \cdot (k \nabla T), \end{aligned} \tag{2}$$

Here, $\rho$ is the fluid density (kg/m³), $u$ is the fluid-velocity vector (m/s), $p$ is the fluid pressure (Pa), $\mu$ is the dynamic viscosity (Pa·s), $C_p$ is the specific heat capacity at constant pressure (J/(kg·K)), and $T$ is the fluid temperature (K).

Computational fluid dynamics (CFD) simulations can evaluate airflow distribution in air-cooled heat sinks, turbulent structures within liquid-cooled microchannels, and two-phase flow boiling heat transfer characteristics. Particularly in research on liquid-cooled microchannels with high heat flux densities, capturing gas-liquid interface evolution using the volume of fluid (VOF) model has become a standard approach for predicting critical heat flux, bubble generation, and flow instability[5]. However, due to the strong nonlinearity of microscale flows, CFD models typically require extensive experimental data for calibration.

## 2.2 Device-Level and Micro/Nanoscale Thermal Simulation

In actual semiconductor devices, heat transport processes are often jointly influenced by intrinsic material phonon properties, interfacial thermal resistance, and device geometry and operating conditions. To address the difficulty of directly applying atomic-scale simulation methods to complex device structures, recent studies have proposed multiscale modeling frameworks for device-level thermal management. By combining phonon heat transport parameters obtained from first-principles calculations and interfacial thermal resistance models with continuum heat conduction equations, these frameworks achieve quantitative predictions of temperature distribution and hotspot behavior in power devices and heterogeneous integration structures. Related work by Xia et al.[6] has been used to analyze the impact of packaging structures, interface engineering, and heat dissipation paths on the thermal reliability of wide-bandgap semiconductor devices, providing important theoretical basis for device structure optimization and thermal design.

As device feature sizes continue to shrink and packaging structures evolve toward three-dimensional and heterogeneous integration, heat transport behavior within semiconductors gradually deviates from macroscopic continuum assumptions, exhibiting significant size effects, interface effects, and non-equilibrium characteristics. In such cases, traditional empirical models based on Fourier's law struggle to accurately characterize real heat transport processes. Therefore, it is necessary to introduce multiscale numerical simulation methods starting from intrinsic material physical mechanisms. Currently, device-level and micro/nanoscale thermal simulations can be mainly categorized into three types: first-principles-based heat transport calculations, mesoscale molecular dynamics (MD) simulations, and data-driven machine learning modeling methods.

### 2.2.1 First-Principles Calculations of Semiconductor Heat Transport and Phonon Properties

Understanding intrinsic heat transport mechanisms in materials is crucial for micro/nanoscale heat transport problems. Traditional macroscopic thermal models often ignore the quantum mechanical nature of heat carriers, making it difficult to accurately predict complex thermal behaviors at the device level. Consequently, a series of first-principles heat transport calculation methods based on density functional theory (DFT) have been widely applied. These methods do not rely on empirical parameters. They obtain thermal conductivity and phonon transport properties by solving the electronic structure and lattice dynamics information of materials. They have been proven to provide reliable physical predictions for nanoscale/microscale materials and device materials. First-principles methods typically obtain lattice constants, phonon dispersion relations, and interatomic force constants from DFT calculations. These results are then input into the phonon Boltzmann transport equation (PBTE) framework to solve for lattice thermal conductivity and frequency-dependent phonon transport behavior, thereby revealing the microscopic mechanisms of phonon scattering and transmission[7,8].

In this system, the phonon PBTE is central to constructing material thermal conductivity models. It describes the transport behavior of heat carriers (mainly phonons) under non-equilibrium conditions by considering factors such as phonon group velocity, phonon lifetime, and phonon-phonon scattering processes. Compared with traditional elasticity theory, the PBTE method can use second- and third-order force constants extracted from DFT for full iterative solutions. This avoids simple relaxation time approximations and improves prediction accuracy. Such methods do not rely on empirical parameters. They obtain lattice thermal conductivity and phonon transport properties by solving the electronic structure and lattice dynamics information of materials, providing reliable intrinsic thermal property inputs for device-level models. Based on this method, many studies have systematically calculated the thermal conductivity of different material systems, such as two-dimensional materials, superlattices, and complex crystal structures. These studies indicate that phonon transport is significantly influenced by material anisotropy, anharmonicity, and phonon-phonon interactions[7,8].

In the context of device-scale thermal management, first-principles phonon transport calculations have been widely used to analyze the intrinsic thermal conductivity and interfacial thermal resistance of semiconductor and two-dimensional material systems. For example, regarding two-dimensional materials, related studies show that the heat transport behavior of materials such as boron monolayers and graphene is mainly controlled by high-frequency phonon branches. This provides important theoretical reference for chip heat dissipation interface engineering using low-dimensional materials[9]. In transition metal dichalcogenide systems, research based on DFT and PBTE reveals the significant suppression of thermal conductivity by nanoscale size effects and strong anharmonicity. This characteristic is particularly critical in power devices and highly integrated chips[10].

Furthermore, first-principles heat transport methods have been systematically applied to research on materials related to silicon-based and wide-bandgap semiconductor devices. This includes size-dependent behavior in nanostructures, heterojunction interfacial thermal resistance, and the potential of two-dimensional materials as thermal interface materials (TIM). Scholars such as Yang et al.[11-18] have conducted extensive pioneering work on phonon transport theory, interfacial thermal conductance regulation, and nanoscale thermal management, providing an important physical foundation for chip thermal design. In addition, combining first-principles methods with high-throughput computing has become an important development direction for screening high/low thermal conductivity materials to serve electronic device thermal management. This approach enables rapid screening of materials with high or low thermal conductivity suitable for device thermal management from large-scale material databases, providing a theoretical basis for microelectronic and thermoelectric device design[19]. Figure 1 presents the high-throughput estimation framework. The high-throughput thermal conductivity estimation process based on elastic properties mainly includes three stages: collecting structural information, calculating phonon properties through micro-volume differentiation, and estimating lattice thermal conductivity by integrating phonon contributions. These correspond to the blue, cyan, and orange parts in Figure 1, respectively[19].

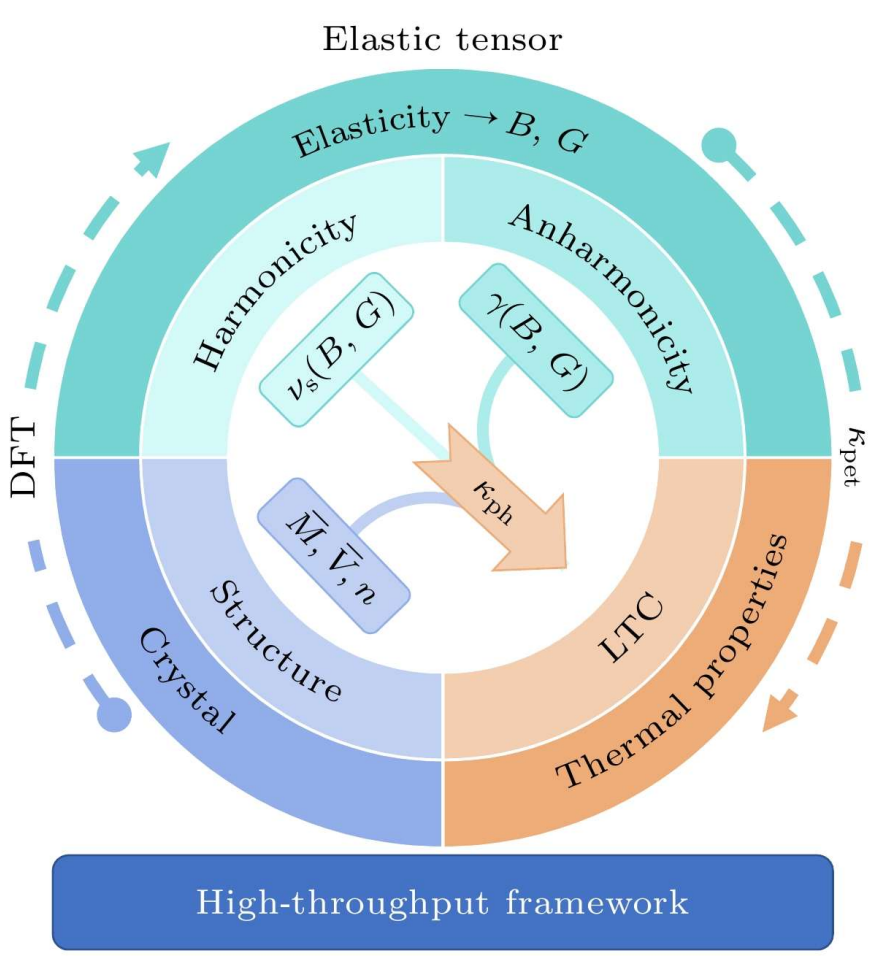

**Fig. 1 A high-throughput estimation framework for lattice thermal conductivity based on elastic properties, the blue section is for collecting basic structural information, the cyan section involves the calculation of elastic properties, including both harmonic and anharmonic properties, the orange section is for estimating the lattice thermal conductivity using the results of elastic property calculations[19].**

In recent years, programs such as ShengBTE[20], Phono3py[21,22], and Alamode[23] have been widely used in first-principles studies of thermal transport. Among these, Alamode demonstrates strong applicability to two-dimensional materials, topological materials, and complex semiconductors. It achieves this through flexible force constant fitting and an efficient framework for solving the phonon Boltzmann transport equation (PBTE), enabling the treatment of complex crystal structures and low-symmetry systems. First-principles methods fundamentally reveal the microscopic origins of thermal transport in materials. They provide reliable intrinsic thermal property inputs for device-level thermal models. In first-principles thermal conductivity calculations, ShengBTE offers a standardized solution. Its workflow clearly separates different computational stages: harmonic force constants are provided entirely by external programs (supporting

Phonopy[21,22] or Quantum Espresso formats), while anharmonic force constants are generated via built-in finite-difference supercell scripts. Finally, the code integrates dielectric properties to apply long-range corrections before calculating thermal conductivity. This design allows researchers to flexibly choose different density functional theory (DFT) backends. As shown in Figure 2, the typical workflow combining Phonopy and VASP[24] has been adopted by multiple research groups[20]. The core algorithmic implementations in the Phonopy and Phono3py codes are illustrated in Figure 3. This method uses a regular grid to divide the parallelepiped region of the reciprocal unit cell into multiple micro-regions of identical shape. The shape of each micro-region is defined by the reciprocal basis vectors $(\boldsymbol{a}_m^*, \boldsymbol{b}_m^*, \boldsymbol{c}_m^*)$. First-principles thermal transport calculations not only reveal the microscopic origins of material thermal conductivity but also provide key intrinsic thermal parameters for device-level thermal management. By accurately calculating lattice thermal conductivity and phonon transport properties, one can predict hot spot formation, heat flux distribution, and thermal resistance effects at material interfaces within device-level models. These methods link microscopic material transport properties to the thermal behavior of nanoscale and microscale devices. Consequently, they provide theoretical guidance for designing novel materials with high or low thermal conductivity and for optimizing device heat dissipation.

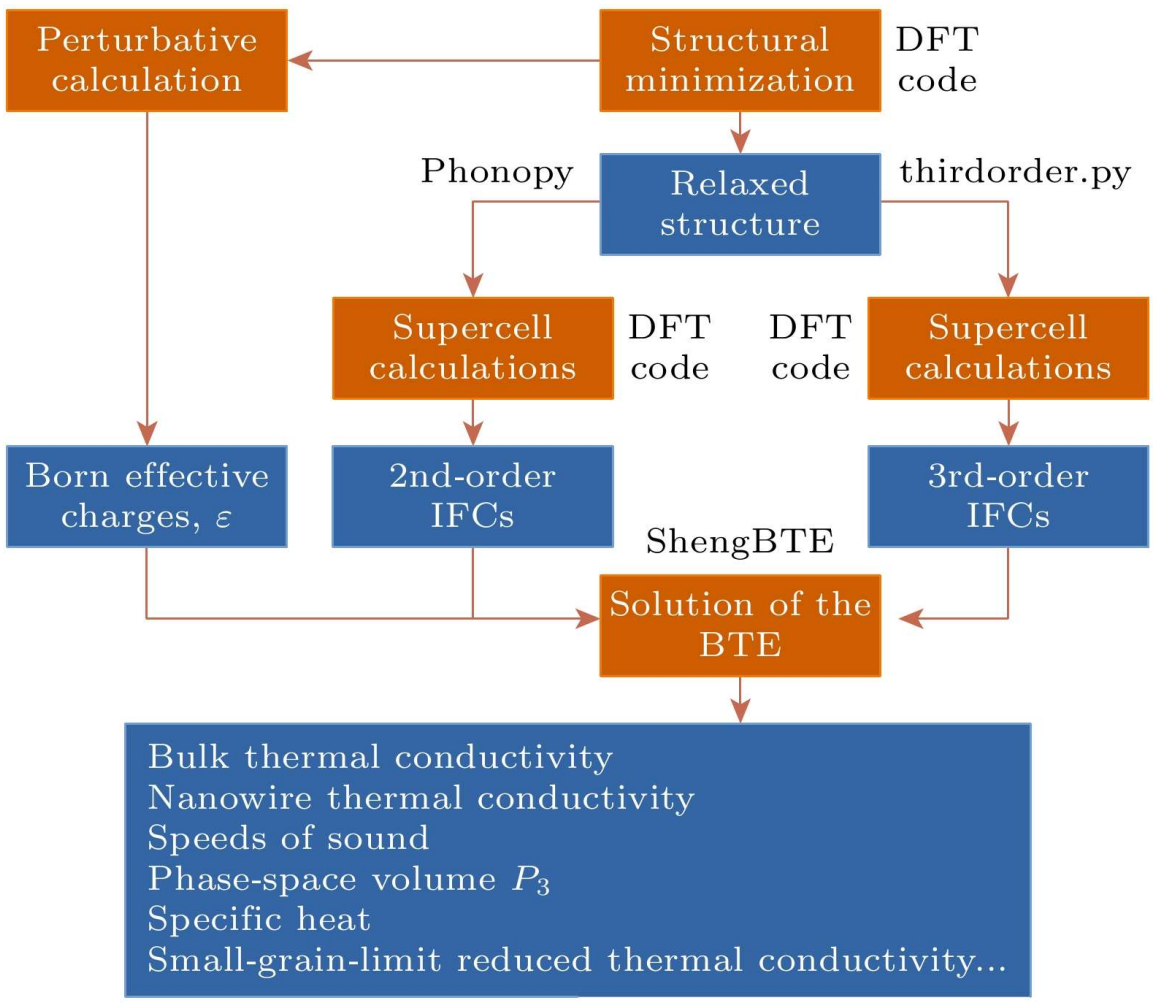

**Fig. 2 The workflow for calculating thermal conductivity by computing force constants using the real-space supercell method, orange boxes represent calculation steps, blue boxes indicate the output results of each step, and computer programs are marked in black text without frames[20].**

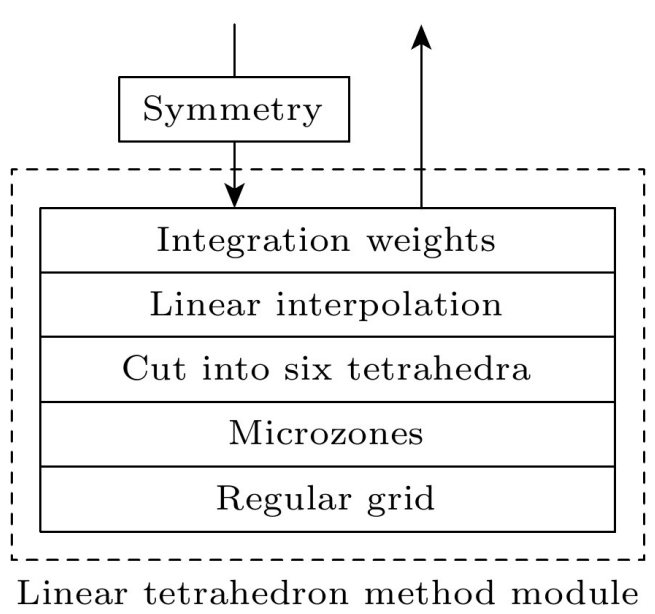

**Fig. 3 The implementation module of the linear tetrahedral method in the Phonopy and Phono3 py codes[21,22].**

Although first-principles methods offer high reliability and atomic-scale physical insights, their computational cost increases rapidly with the number of atoms. This makes direct application to real devices or packaging structures challenging. Consequently, these methods primarily serve to provide intrinsic material parameters and physical understanding, supplying reliable inputs for subsequent mesoscopic and continuum models.

### 2.2.2 Molecular-Dynamics Simulation of Mesoscale Semiconductor Heat Transport

In micro- and nanoscale heat transport simulations, classical molecular dynamics (MD) has become a crucial tool for understanding and predicting the thermal conductivity, thermal boundary resistance (TBR), and non-equilibrium heat flow behavior of complex semiconductor materials. This advantage stems from its ability to accurately describe atomic motion at the atomic scale. MD simulations drive the dynamic evolution of atomic systems over time based on Newton's equations of motion. By constructing appropriate interatomic potential functions to represent atomic forces, MD can simultaneously capture the effects of phonon-phonon scattering, interface scattering, and size effects on heat transport. These effects are difficult to incorporate into traditional continuum models. The core advantage of MD lies in its independence from macroscopic continuum assumptions. It reveals the fundamental mechanisms of heat flow in systems where the scale approaches the phonon mean free path. This approach has direct application value for complex semiconductor systems commonly found in modern device structures, such as nanowires, thin films, and heterogeneous interfaces[25]. Active learning strategies combined with high-throughput MD simulations have been employed to screen polymer blends with high thermal conductivity. These studies also investigate the underlying mechanisms by which certain blends exhibit higher thermal conductivity than their constituent single polymers. The specific implementation is illustrated in Figure 4[25].

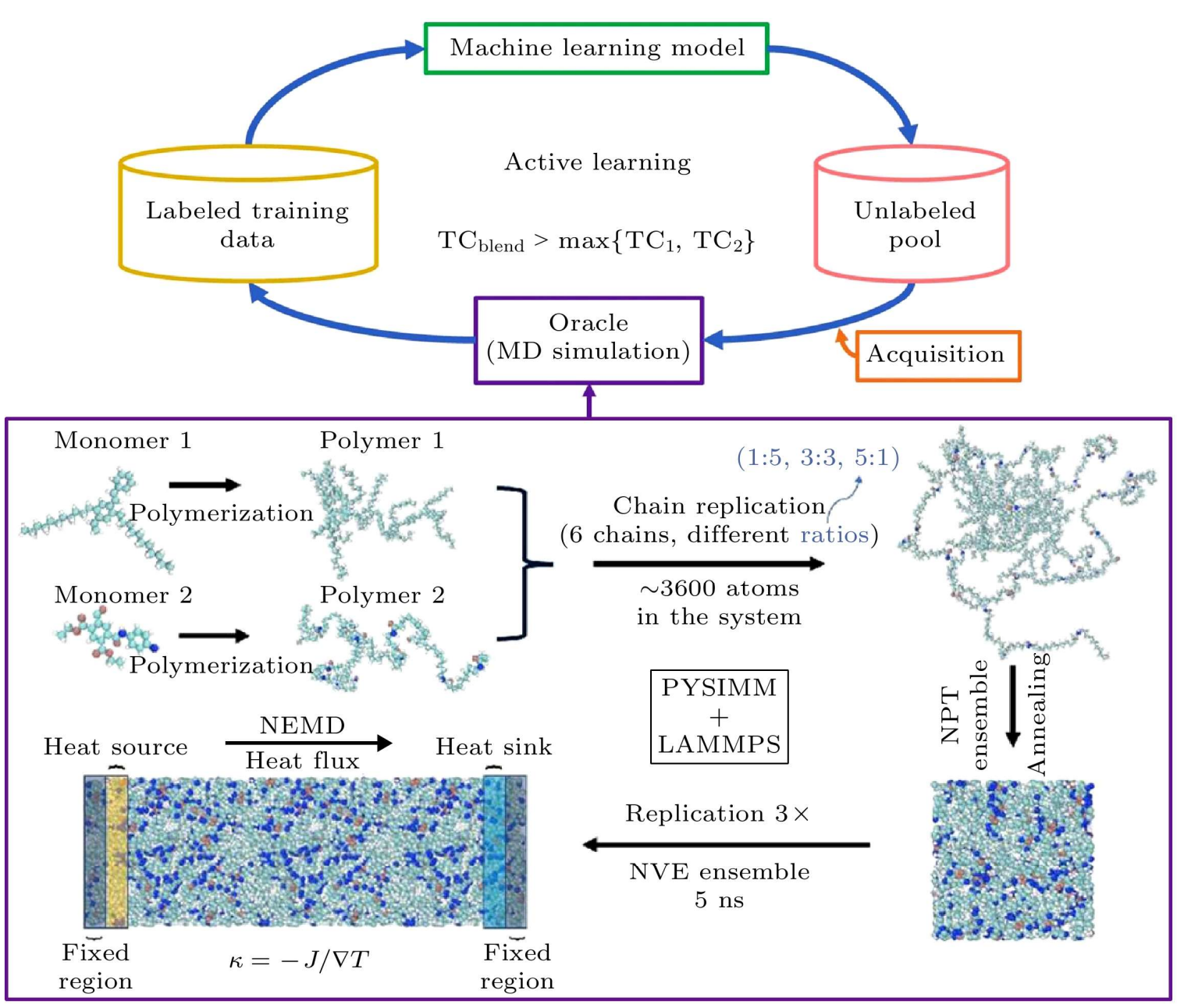

**Fig. 4 Schematic diagram of the screening framework for high thermal conductivity polymer blends based on high-throughput MD and active learning strategies[25].**

In practical implementation, MD simulations typically employ two primary methods to calculate thermal conductivity: equilibrium molecular dynamics (EMD) and non-equilibrium molecular dynamics (NEMD). The EMD method relies on the Green-Kubo relation. It obtains the thermal conductivity of the system through time integration of the heat flux autocorrelation function. This approach is suitable for statistical analysis of heat transport under thermal equilibrium fluctuations. Its advantage lies in not requiring an artificially imposed external temperature gradient. This avoids interference from nonlinear heat transfer effects. Furthermore, it is easy to implement under periodic boundary conditions. It is particularly suitable for describing the intrinsic thermal conductivity of bulk materials. However, EMD demands high statistical sampling. Long simulation times are necessary to ensure sufficient convergence of the heat flux correlation function. Meanwhile, in low-dimensional materials or systems with strong size effects, the results may be influenced by finite-size effects. In contrast, the NEMD method establishes a steady-state temperature gradient by applying heat and cold sources at both ends of the system. It directly calculates thermal conductivity based on Fourier's law. This method can intuitively characterize non-equilibrium heat transfer processes. It is especially suitable for studying transport problems dominated by size effects and interface scattering. Its advantages include an intuitive physical picture. Temperature distribution and heat flux can be directly observed. It is particularly suitable for investigating thermal transport behavior in interface thermal resistance, size effects, and heterogeneous systems. However, NEMD often requires large system sizes to avoid unphysical perturbations of the temperature field by the heat/cold source regions. In nanoscale systems, it may be affected by non-local heat transfer or ballistic transport effects. This can lead to systematic deviations in the thermal conductivity extracted via Fourier's law.

When quantitatively predicting TBR, MD results often depend heavily on the employed empirical potential functions. Their accuracy in describing cross-interface bonding, force constants, and anharmonic interactions is critical. Therefore, certain uncertainties remain[26]. Recent studies indicate that if the potential function fails to accurately reproduce interface vibration spectra and bonding characteristics obtained from experiments or first-principles calculations, MD may produce significant deviations. These results may differ from actual TBR values. This issue is particularly prominent in complex heterojunctions. Feng et al.[27-29] systematically compared MD, atomistic Green's function, and first-principles methods. They emphasized the importance of introducing higher-precision atomic-scale models and multiscale computational frameworks in interface heat transport research.

Nevertheless, MD simulations retain unique advantages in revealing phonon cross-interface scattering and transmission mechanisms in semiconductor heterostructures. By constructing models with different interface roughness, atomic mixing degrees, or chemical bonding modes using NEMD, and by statistically analyzing heat flux and temperature jumps, one can systematically analyze the impact trends of interface structural changes on TBR. This allows for testing the applicability of classical theories such as the acoustic mismatch model and the diffuse mismatch model.

Such atomic-scale simulations not only reveal the physical mechanisms of phonon scattering and transmission at interfaces but also provide qualitative or even semi-quantitative guidance for interface engineering design[25,30,31]. For example, in MD heat transport simulations of alternating dielectric thin films and superlattice structures, studies show that interface scattering becomes the main regulating factor for heat flow when interface density increases or layer thickness decreases. This effectively suppresses overall thermal conductivity. This result holds significant reference value for understanding interlayer thermal resistance contributions in three-dimensional integrated chips[32].

Furthermore, MD simulations are widely used to study the heat transport properties of semiconductor materials, such as Si, GaN, and SiC. For high-power electronic devices, such as GaN high electron mobility transistors (HEMTs), MD simulations can systematically analyze the impact of crystal structure, defects, and interfaces on phonon scattering and heat transport. This provides a precise numerical framework for near-junction thermal management[33]. Studies indicate that in amorphous silicon-carbon thin films, thermal conductivity exhibits obvious temperature and size dependence as temperature rises or system size changes. This mainly stems from changes in phonon mean free path. In comparative studies of silicon nanowires and GaN nanostructures, different MD heat transport methods (such as EMD and NEMD) show differences in evaluating thermal conductivity and phonon density of states spectra. This highlights the importance of using multiple simulation methods for collaborative analysis[34].
Although MD simulations play an important role in revealing the physical mechanisms of micro- and nanoscale heat transport, the accuracy of their results is still limited by factors such as the accuracy of the employed classical potential functions, system scale, and boundary conditions. Therefore, to improve the predictive capability of MD simulations, current research is gradually combining MD with machine learning potential functions and first-principles parameters. The aim is to achieve higher-precision large-scale heat transport simulations. This direction will be discussed further in subsequent sections.

### 2.2.3 Machine-Learning Simulation of Semiconductor Heat Transport for Device and Structural Optimization

As chip structures become more complex and the number of design parameters grows, conventional heat-transport simulations face a fundamental trade-off between computational efficiency and accuracy. In recent years, machine-learning interatomic potentials and high-performance-computing methods have expanded the capabilities of MD simulation. By combining accurate potentials with efficient computational frameworks, these approaches improve both the reliability and scalability of predictions for heat transport in complex materials. DeePMD[35-37] (deep-potential molecular dynamics) and GPUMD[38] (graphics-processing-unit molecular dynamics) are two representative approaches that have had a substantial impact on semiconductor heat-transport simulation.
The DeePMD method learns interatomic interaction potentials from high-quality first-principles data through deep neural networks. It uses these as machine learning potentials for large-scale MD simulations. This effectively bridges the gap in accuracy and efficiency between traditional empirical potentials and first-principles methods. As an important implementation platform for the DeePMD method, DeePMD-kit makes a core contribution by supporting the efficient training of high-dimensional neural network potentials from first-principles data. It embeds the obtained potential functions into large-scale molecular dynamics simulations. This enables fine characterization of many-body interactions and anharmonic phonon effects. Such methods typically adopt local atomic environment descriptors. During the training process, they jointly fit energy, forces, and stress tensors. This allows for reliable prediction of physical quantities closely related to heat transport, such as phonon spectra, interface scattering, and defect-induced thermal resistance[36]. The software architecture and workflow are illustrated in Figure 5.

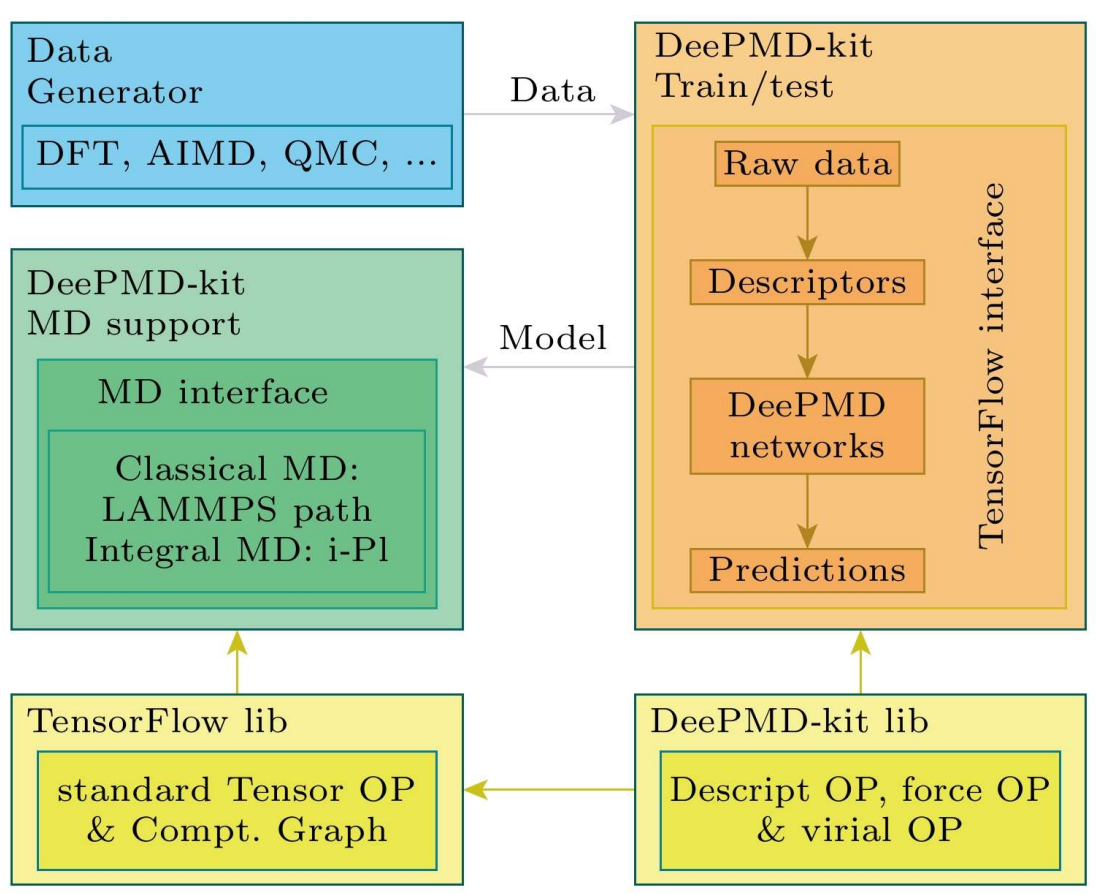

**Fig. 5 Architecture diagram and workflow of DeePMD-kit[35–37].**

GPUMD[38] represents a technical approach that deeply integrates machine learning potentials with GPU-accelerated computing. Its integrated neuroevolution potential (NEP) model is specifically optimized for calculating heat flux and stress tensors. This optimization enables million-atom-scale thermal transport simulations on a single GPU platform while maintaining accuracy comparable to first-principles methods. Related studies indicate that this framework can effectively quantify the effects of grain boundaries, interfaces, and structural disorder on thermal conductivity and interfacial thermal resistance. It also supports decomposition analysis of the contributions from different phonon modes to transport. The complete "training-application" workflow is shown in Figure 6.

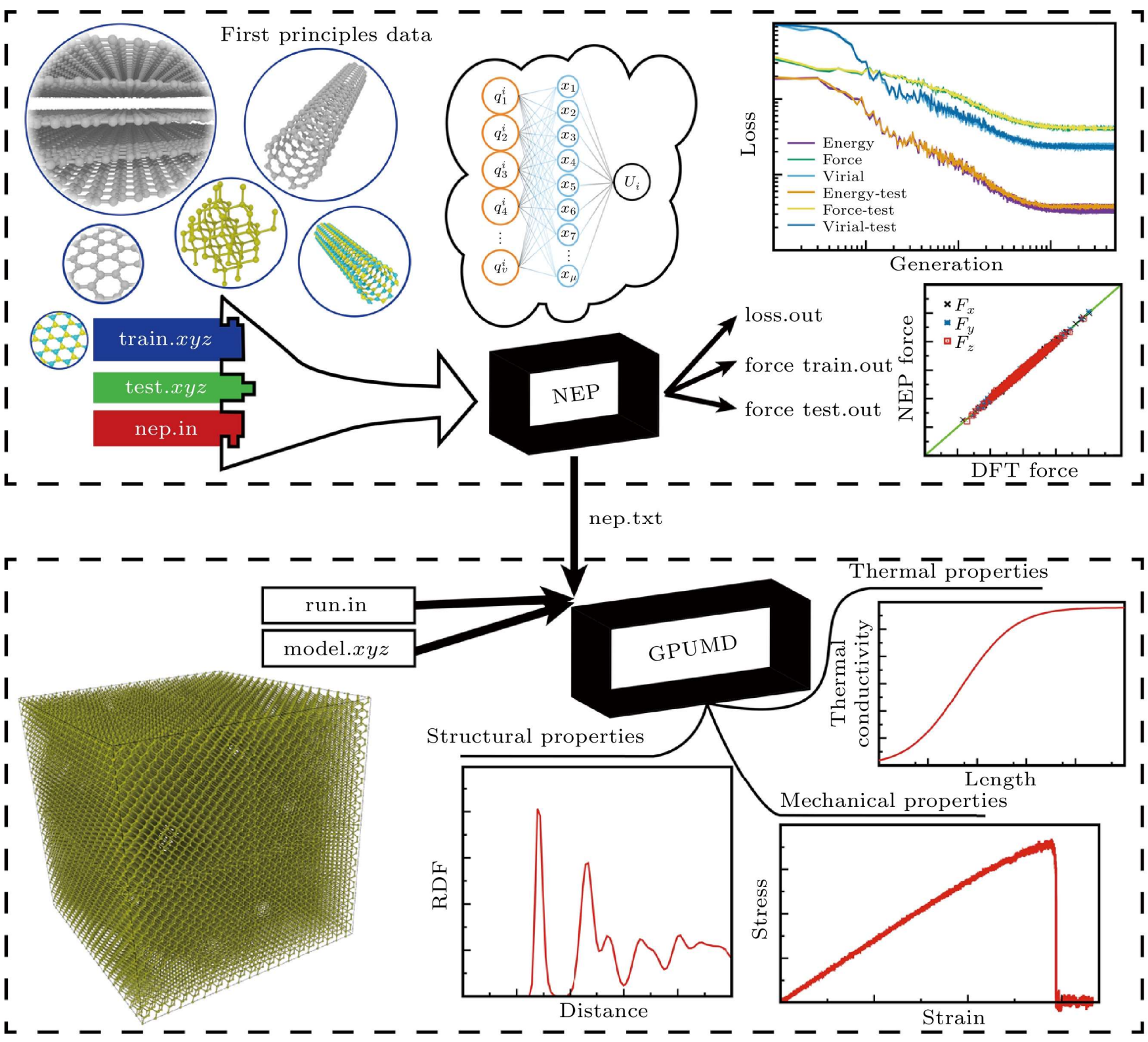

Fig. 6 GPUMD workflow, the GPUMD software package consists of two executable programs of nep and gpumd, represented by two black boxes respectively, among them, the nep program is used to train the NEP potential function model; the gpumd program uses the trained potential function to conduct atomic-scale simulations[38].

By integrating DeePMD with GPUMD, researchers can efficiently conduct large-scale molecular dynamics (MD) simulations of thermal transport properties. For instance, using the neuroevolution potential (NEP) machine learning potential and the GPUMD platform, previous studies have successfully achieved near-first-principles accuracy in simulating the thermal transport behavior of polycrystalline graphene with millions of atoms on desktop GPUs. These studies quantified the impact of grain boundaries on thermal conductivity and resolved the transport contributions of different phonon modes, providing important physical insights into thermal management issues in low-dimensional materials[39]. Furthermore, reviews based on GPUMD explicitly state that machine learning potential methods not only improve computational efficiency in thermal transport simulations but also enable the accurate capture of physical phenomena such as anharmonic effects, spatial disorder, and interface scattering in complex materials within the MD simulation framework[40].

Machine learning-driven MD simulations have also promoted the development of multiscale thermal modeling. On one hand, high-precision machine learning potentials can automatically learn complex potential energy surfaces from first-principles data, thereby improving the accuracy of thermal transport parameters. On the other hand, high-performance MD platforms leverage GPU architectures to provide the computational power for large-scale simulations, making the thermal behavior of semiconductor systems containing numerous defects, interfaces, and heterogeneous structures computationally tractable. This simulation strategy, built on the combination of "machine learning potentials + high-performance MD software," is gradually becoming a key bridge between thermal transport research and device-level thermal design. It holds broad application prospects in complex conditions such as 3D integration, heterogeneous packaging, and thermo-mechanical coupling.

## 2.3 Multiphysics-Coupled Simulation

As device dimensions shrink and operating frequencies increase, thermal effects become tightly coupled to electric fields, carrier mobility, and mechanical stress. Coupled electrothermal models simultaneously solve the electric-potential, carrier-transport, and heat-conduction equations to capture Joule heating, mobility degradation, and the feedback of temperature on electrical performance.

This electrothermal coupling effect not only determines the transient and steady-state performance of devices but also directly affects their long-term reliability and lifetime metrics. In device reliability assessment, common lifetime characterization metrics include mean time to failure (MTTF) and failure mechanisms accelerated by thermal effects, such as electromigration (EM) in metal interconnects and negative bias temperature instability (NBTI) related to gate dielectrics. These failure mechanisms are highly sensitive to junction temperature, with their degradation rates typically increasing exponentially with temperature. For example, in GaN HEMTs, high-power-density pulses cause rapid temperature rises in the gate region, leading to a decrease in carrier mobility by approximately 20%-30%. This not only weakens the transient output performance of the device but also accelerates temperature-related degradation processes, thereby shortening the device MTTF[41].

During semiconductor-device operation, heat transport is closely coupled to electrical transport and the operating state, so a heat-conduction model alone cannot accurately describe the internal temperature and power-dissipation distributions. To address this limitation, Li et al.[42] developed coupled electrothermal simulation methods that integrate carrier transport, Joule heating, and heat diffusion in a multiphysics framework. The framework was used to analyze temperature rise and thermal reliability in wide-bandgap semiconductor devices under high power density and complex cooling conditions. These studies reveal how multiphysics coupling affects device performance and stability through device structure, interface engineering, and co-optimization of the cooling system, providing guidance for thermal management and system-level optimization of high-power electronic devices.

Thermo-mechanical coupling models are particularly important at the package level. Large differences among the coefficients of thermal expansion of silicon, copper, and packaging epoxy can cause temperature gradients to generate stress concentrations of hundreds of megapascals around through-silicon vias (TSVs), making thermal stress a major contributor to metal fatigue and interconnect failure[43]. Multiphysics-coupled simulation has therefore become an essential tool for reliability analysis in advanced packaging.

## 2.4 Recent Advances in AI-Accelerated Thermal Simulation

As chip thermal problems become more complex, deep-learning surrogate models are increasingly being incorporated into thermal-analysis workflows. By learning mappings from layouts and material parameters to temperature fields, deep neural networks can reduce calculations that take hours to millisecond-scale inference while retaining high accuracy. For example, Gabourie et al.[44] proposed a U-Net-based model for predicting effective thermal conductivity in complex package designs. Within the structural-parameter range represented by the training and test data, the model achieved approximately 95% predictive accuracy relative to high-fidelity numerical simulations and a speedup of more than three orders of magnitude. Such AI-accelerated simulation makes real-time chip thermal analysis and runtime temperature prediction possible and provides a foundation for intelligent thermal design.

Building on this, researchers have further introduced models such as convolutional neural networks (CNNs), graph neural networks (GNNs), and physics-informed neural networks (PINNs) into multiscale heat conduction problems in recent years. These models are used to characterize temperature evolution behaviors under conditions of complex geometric structures, non-uniform material distributions, and multi-heat-source coupling[45-49]. Among these, GNNs demonstrate strong generalization capabilities in handling irregular packaging morphologies and hierarchical interconnects by representing chip layouts and interconnect structures as topological graphs[45,49]. Meanwhile, PINNs explicitly embed heat conduction governing equations into the loss function, ensuring consistency of network predictions under physical constraints. This approach improves model stability and interpretability while reducing the number of training samples required[47,48].

Hybrid data-driven and physics-based modeling frameworks are also becoming an active area of research. In these frameworks, finite element or finite volume simulations generate a limited number of high-fidelity samples as supervisory data, and deep-learning models then construct fast thermal surrogates for large-scale parameter sweeps and design-space exploration[50-52]. These methods are particularly well suited to advanced packaging and heterogeneous integration, including complex vertical heat conduction in chiplet architectures, 2.5D/3D stacked chips, and TSV interconnect systems, and can substantially reduce the computational cost of thermal design.

It is worth noting that the reliability and generalization capability of AI models in engineering applications remain key challenges in current research[53]. On one hand, model performance highly depends on the distribution of training data; when geometric structures or material parameters exceed the coverage of the sample set, prediction accuracy may degrade significantly. On the other hand, providing uncertainty quantification and error assessment mechanisms while maintaining the advantage of inference speed is an important prerequisite for promoting their adoption in actual chip design workflows[54,55]. Therefore, recent works have begun to combine Bayesian neural networks, ensemble learning, and active learning strategies to evaluate model prediction confidence intervals and continuously expand training datasets through adaptive sampling[56]. Table 1 and Table 2 list the comparisons of different numerical simulation methods and the mapping relationships between scales and methods, respectively.

**Table 1.** The main characteristics of typical numerical simulation methods in chip thermal management.

| Field of application | Representative method | Main contribution | References |
| --- | --- | --- | --- |
| Macro and chip level thermal simulation | Equivalent thermal circuit model | The chip and package are simplified as a thermal resistance-heat capacity network to analyze the overall temperature distribution and thermal response. | [3] |
| Device level thermal simulation | FEM | Fine description of temperature field and material inhomogeneity in device | [4] |
| Thermal-fluid coupling simulation | CFD | The flow and heat transfer behavior of cooling medium is studied, and the heat dissipation problem of high heat flux can be analyzed. | [5] |
| Microscopic heat transport simulation | Phonon transport model | Non-Fourier heat conduction and size effect can be described to reveal the mechanism of micro-nano scale heat transport. | [7,8] |
| Atomic scale simulation | MD | Study TBR and atomic-level heat conduction behavior, with clear physical mechanism, suitable for interface problem | [25] |
| Intelligent Accelerated Simulation | AI accelerated thermal simulation | Data-driven prediction of temperature field and thermal response, fast calculation speed; suitable for large parameter space | [44] |

**Table 2.** Scale-method mapping table.

| Scale hierarchy | Length scale | Time scale | Representative method | Suitable material | Primary output | Computational cost | Typical sources of error |
| --- | --- | --- | --- | --- | --- | --- | --- |
| System level | cm—mm | ms—s | Equivalent thermal circuit RC | SoC, Chiplet Initial Evaluation | T, hot spot prediction | Very low | The power map is simplified and the boundary conditions of the are coarse. |
| Package Level | mm—μm | ms—s | FEM/FVM | TSV, HBM Stacked Package | T-field, heat flow, stress | Medium-high | Insufficient mesh, unknown TBR |
| Thermal-fluid coupling | cm—μm | ms—s | CFD (VOF two-phase) | Microchannel liquid cooling, jet cooling | Heat transfer coefficient, pressure drop | Extremely high | Turbulence Model, Phase Change Parameter Dependence Experiment |
| Mesoscopic non-Fourier | mm—nm | ns—μs | Phonon BTE/PBTE | Nanofilm, superlattice | Kappa-spectrum, non-equilibrium heat flow | High | Scattering mechanism assumption, material parameters |
| Atomic scale interface | nm—Å | ps—ns | MD/NEMD | Heterointerface TBR, defect | TBR, phonon scattering | Extremely high | Potential function accuracy, finite size effect |
| Data Driven Agent | Arbitrary | Real-time prediction | AI surrogate/ PINN | Rapid thermal prediction and optimization | Temperature field, power inversion | Training High/ Reasoning Low | Generalization difference outside data distribution |

# 3 Experimental Techniques

Compared with numerical simulation, experimental measurement plays a critical role in parameter calibration, model correction, and validation of physical failure mechanisms in chip thermal design. Because internal chip temperature fields are characterized by high heat fluxes, strong transients, and submicrometer features, experimental techniques must combine high spatial and temporal resolution with accurate thermophysical-property measurement. This section reviews temperature-field measurement, thermal-boundary-resistance and thermophysical-property characterization, and joint simulation-experiment validation.

## 3.1 Temperature-Field Measurement Techniques

The principal temperature-field measurement techniques include infrared thermography (IRT), thermoreflectance (TR), Raman thermometry, and embedded sensors. Because of its simple operation and large field of view, infrared imaging is the most widely used method for package-level temperature measurement. IRT derives two-dimensional surface-temperature maps from the relation between a material's emissivity and temperature. Its spatial resolution is limited by the infrared wavelength, however, so it cannot readily resolve nanoscale devices or submicrometer hotspots. In addition, the high reflectivity of metal interconnects requires emissivity calibration, often using a blackbody coating, to obtain reliable measurements[57]. IRT is therefore used mainly to evaluate package-level heat-sink performance and validate convective heat-transfer experiments.

Thermoreflectance is an optical technique that exploits the temperature dependence of a material's reflectivity. By measuring changes in the intensity or phase of reflected light, it can infer the surface-temperature field and thermophysical parameters such as thermal diffusivity and thermal resistance:

$$\frac{\Delta R}{R} = C\Delta T, \tag{3}$$

Here, $\Delta R$ is the change in reflectivity and $R$ is the reflectivity at a reference temperature. The normalized change $\Delta R/R$ is typically assumed to vary linearly with the temperature change $\Delta T$, where $C$ is the thermoreflectance coefficient ($K^{-1}$), usually of order $10^{-5}$-$10^{-4}$ $K^{-1}$. This approximation is valid for small temperature perturbations[58]. The spatial resolution of TR imaging is generally set by the laser spot size and can reach the micrometer scale, whereas its temporal resolution is limited by the laser-pulse width and can reach the picosecond-to-nanosecond range[59]. TR can measure temperatures at critical locations such as CMOS gates, metal interconnects, and the drain region of a GaN HEMT, and is among the most widely used submicrometer temperature-imaging methods. Raman thermometry is a noncontact, high-spatial-resolution technique based on the temperature dependence of Raman-peak position, intensity, and width. Raman scattering is excited in the specimen, the scattered-light spectrum is collected, and the local temperature is inferred from a calibrated relation between temperature and the Raman response. Figure 7 shows a schematic micro-Raman system[60]. Because the Raman shift is related to phonon energy, increasing temperature generally redshifts the Raman peak; calibration of peak position against temperature enables measurements with approximately 1 μm spatial resolution. Raman thermometry can also provide stress information because thermal stress changes the peak position. It has therefore been used in 3D ICs to measure both temperature and thermal-stress fields near TSVs and to validate thermo-mechanical coupling models[61]. Its weak signal and long scanning time, however, limit large-area and long-duration measurements. Temperature and stress also affect the Raman peak simultaneously, so their separation relies on calibrated material parameters and model assumptions. Laser self-heating, spectral resolution, and peak-fitting errors introduce further uncertainty, especially at micro- and nanoscales and in multimaterial structures. Raman methods are therefore best suited to mapping local relative temperature and stress and remain limited for high-accuracy absolute measurement over large thermal fields. Embedded temperature sensors, including ring oscillators, thermopiles, and TSV-based probes, can directly measure internal chip temperatures in advanced packages. The frequency of a ring oscillator varies monotonically with temperature:

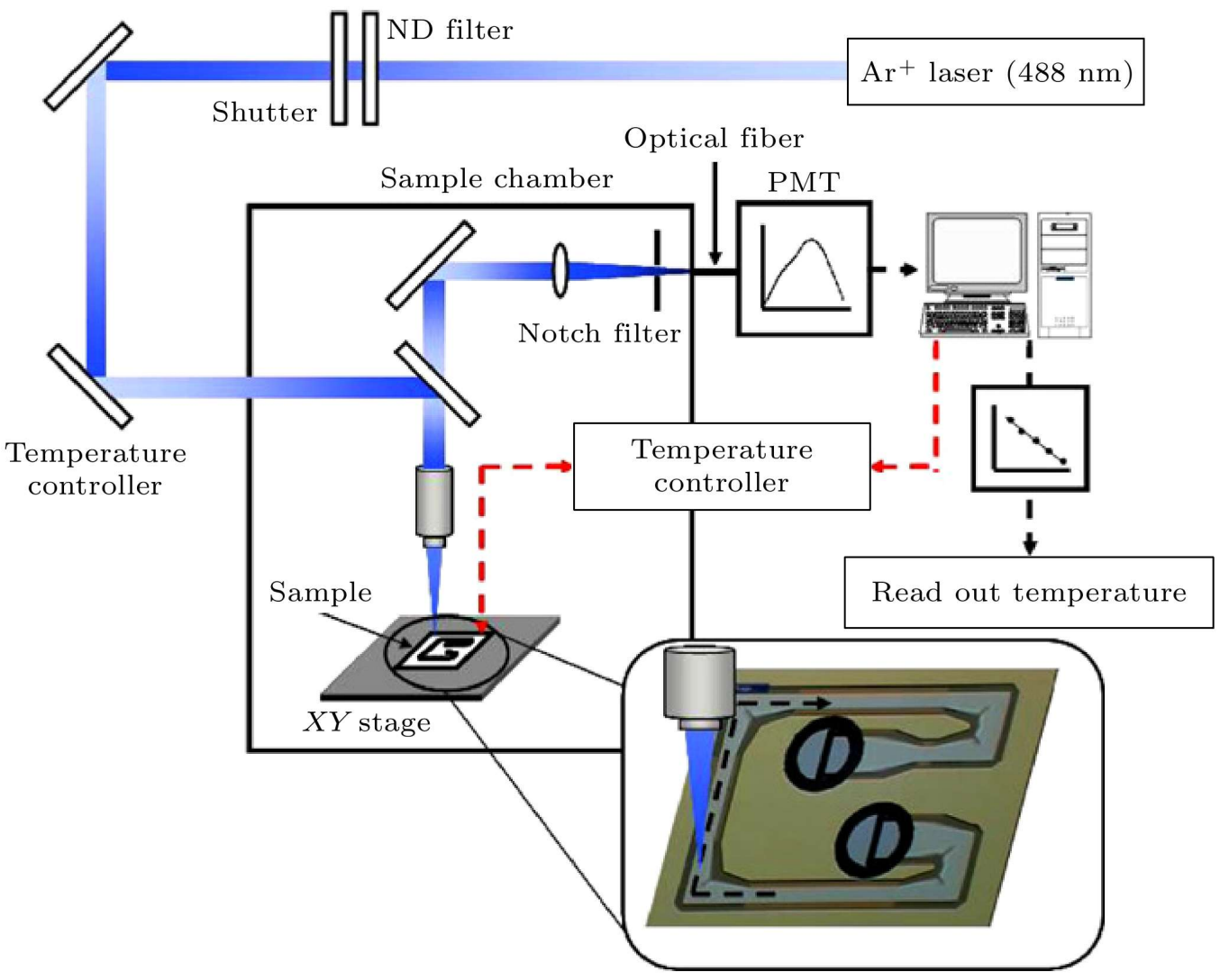

**Fig. 7 Micro-Raman spectroscopy system[60].**

**Table 3.** Comparison of the characteristics of different experimental temperature measurement methods in the research of chip thermal management.

| Measurement scale | Experimental method | Main contribution | References |
|---|---|---|---|
| Package Level/Chip Surface | IRT | Two-dimensional temperature distribution on device surface, non-contact measurement, simple operation, large field of view | [57] |
| Micrometer scale | TR | Local temperature and transient thermal response with high spatial and temporal resolution | [59] |
| Micrometer/submicrometer scale | Raman spectroscopy thermometry | Local temperature and thermal stress distribution, temperature and stress information can be obtained at the same time | [61] |
| Device internal/chip level | Embedded temperature sensor | Chip internal real-time temperature, in-situ measurement, high real-time performance | [62] |

$$f(T) \propto \mu(T), \tag{4}$$

Because the mobility $\mu(T)$ decreases with increasing temperature, ring oscillators can provide real-time temperature monitoring. Embedded sensors enable distributed on-chip temperature sampling and are important for validating power maps and heat-diffusion paths in numerical simulations[62]. Their placement, however, consumes chip area and can perturb local heat flow. Table 3 compares the experimental temperature-measurement methods used in chip thermal-management research.

## 3.2 Measurement of Thermophysical Properties and Thermal Boundary Resistance

Material thermal conductivity, thin-film heat capacity, and thermal boundary resistance (TBR) are critical input parameters affecting simulation accuracy. The primary methods include the $3\omega$ method, time-domain thermoreflectance (TDTR), frequency-domain thermoreflectance (FDTR), and microchannel cooling experimental platforms.

### 3.2.1 The $3\omega$ Method

The 3$\omega$ method is a key experimental technique for measuring thermal properties. It operates by heating with alternating current and measuring the temperature response on the sample surface. This technique is widely used in thin films, nanomaterials, microelectronic devices, and semiconductors. By adjusting the modulation frequency of the heating current (i.e., the frequency of the periodic heating signal, typically in the Hz to kHz range), the method characterizes heat transport behavior across different thermal diffusion lengths and time scales. Consequently, it offers unique advantages for measuring microstructures and TBR[63]. In theoretical analysis, the 3$\omega$ method typically relies on semi-infinite body or multilayer film heat diffusion models. Its applicability depends on the relationship between the thermal diffusion length, sample geometric dimensions, and modulation frequency.

The 3$\omega$ method is based on the coupling between heat diffusion and electrical resistance, as illustrated in Figure 8. A metal resistance line, typically aluminum, platinum, or molybdenum, is fabricated on the sample surface and driven by an alternating current at angular frequency $\omega$. Joule heating produces a temperature oscillation in the line and a corresponding resistance oscillation. The analysis generally assumes that the line is much longer than it is wide or thick, that heat flows primarily normal to the sample surface, and that lateral heat spreading, air convection, and radiation are negligible; heat transport can then be approximated as one-dimensional. Because Joule heating varies at 2ω, the temperature and resistance also contain a 2ω component. Multiplication of this resistance oscillation by the drive current produces a voltage component at 3ω. If the surface temperature response is periodic, the applied current can be written as[63]

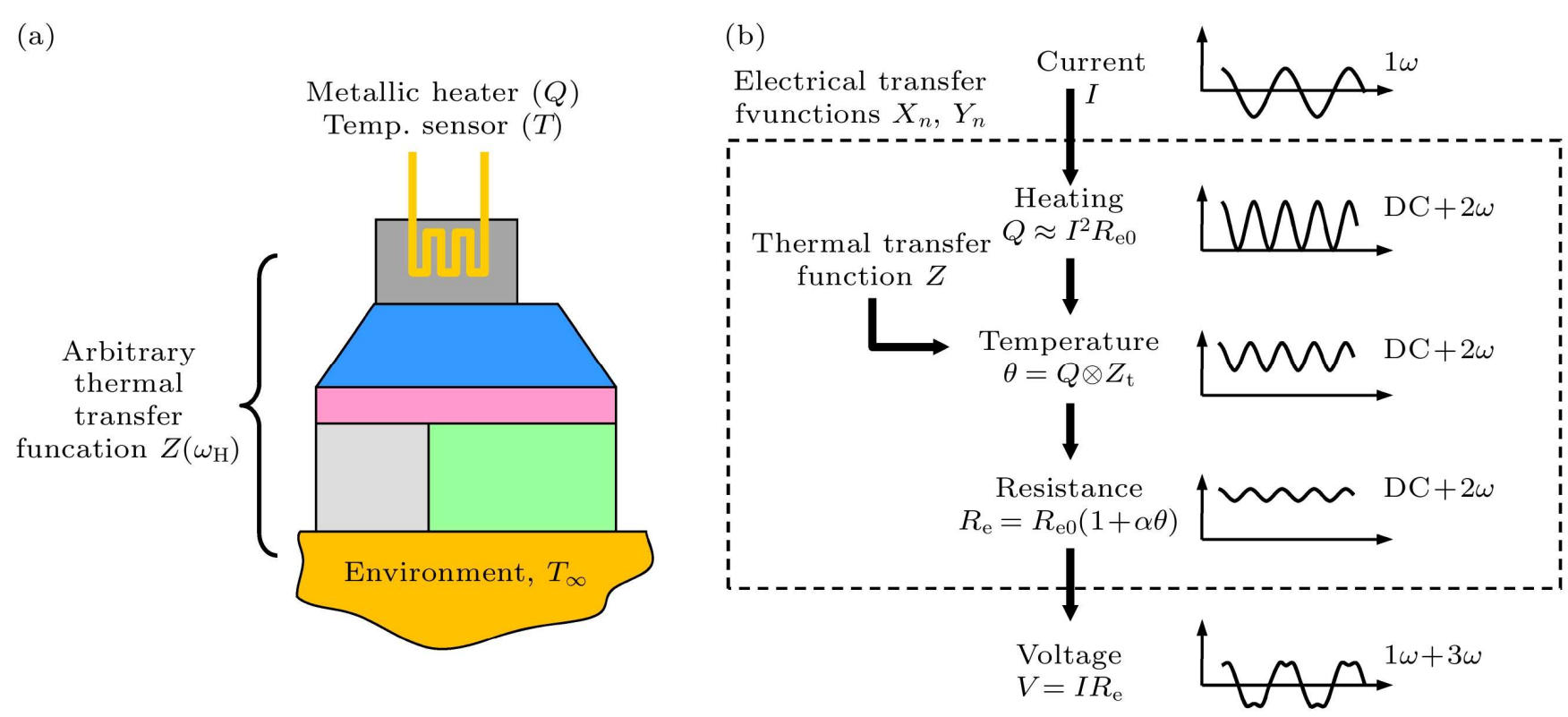

**Fig. 8 (a) A general system for measuring transfer functions using the 3$\omega$ method; (b) relationship between sinusoidal current and voltage, as well as the heat transfer function[64].**

$$I(t) = I_0 \sin(\omega t), \tag{5}$$

Here, $I_0$ denotes the current amplitude, and $\omega$ represents the current frequency. As the current passing through the resistive wire generates heat, the surface temperature of the sample undergoes periodic fluctuations. Assuming that the dominant frequency component of the temperature variation is $2\omega$, the change in resistance $\Delta R(t)$ can be expressed as[63]

$$\Delta R(t) = \alpha \Delta T(t), \tag{6}$$

Here, $\alpha$ denotes the temperature coefficient of resistance. Owing to the relationship between current and resistance, the voltage signal contains a frequency component at $3\omega$. By measuring the $3\omega$ component of the voltage signal, $V_{3\omega}$, one can extract the thermal properties of the material. Under steady-state or periodic heating conditions, $V_{3\omega}$ is typically expressed as a complex number[63]

$$V_{3\omega} = |V_{3\omega}| \mathrm{e}^{\mathrm{i}\phi_{3\omega}}, \tag{7}$$

Here, $|V_{3\omega}|$ denotes the amplitude of the $3\omega$ voltage, and $\phi_{3\omega}$ represents its phase. The phase reflects the phase lag of the temperature response relative to the excitation current. Since metal resistance varies with temperature, measuring the amplitude and phase of the $3\omega$ voltage allows for the inversion of sample temperature changes. According to the heat diffusion equation, the relationship between the temperature response and the frequency $\omega$ is described by the thermal diffusivity. Specifically, the relationship between the temperature response $\Delta T_{2\omega}$ and the frequency $\omega$ can be expressed as follows[63]:

$$\Delta T_{2\omega} = (1/\sqrt{\omega}) \cdot \exp(-\sqrt{\omega/D}), \tag{8}$$

Here, $D$ denotes the thermal diffusivity. By varying the current frequency $\omega$ and measuring the voltage signals at different frequencies, one can obtain the temperature response as a function of frequency. This allows for the determination of the thermal diffusivity and thermal conductivity of the material. The $3\omega$ method is particularly effective for thin films or multilayer structures. It enables accurate measurement of the thermal conductivity of thin-film materials and the thermal boundary resistance (TBR) between layers. TBR is critical for thermal management in microelectronic devices. By measuring temperature responses at various frequencies and incorporating the geometric structure of the thin film, the $3\omega$ method provides a common experimental approach for the quantitative analysis of TBR.

As research shifts from continuous films to low-dimensional nanostructures, the conventional $3\omega$ method faces limitations in spatial resolution and sample compatibility. Electrical-heating and temperature-sensing schemes developed for micro- and nanoscale systems now enable high-precision measurements of individual nanowires. In one approach, a focused electron beam provides a localized heat source at selected positions along a semiconductor nanowire, while sensitive thermometers on suspended microbridge sensors measure the resulting temperature rise. This makes it possible to map local thermal conductivity along a single nanowire with nanoscale spatial resolution. The method has been used to study how defect engineering and ion irradiation alter heat transport and provides an important tool for probing nonuniform transport in low-dimensional structures[65].

### 3.2.2 TDTR and FDTR

Time-domain thermoreflectance (TDTR) and frequency-domain thermoreflectance (FDTR) are two non-contact thermal characterization techniques widely adopted in recent years for studying micro- and nanoscale heat transport. Cahill[66] systematically developed the TDTR method by introducing femtosecond pump-probe laser technology into thermal characterization. This approach enables the generation of heat flux via pump excitation and the monitoring of time-dependent changes in probe reflectivity. Consequently, it allows for the systematic measurement of thermal properties such as thermal conductivity, thermal diffusivity, and interfacial thermal conductance. Early work by Cahill[66] established TDTR as a routine measurement technique with high temporal resolution. It demonstrated broad applicability across various material systems, including thin films, superlattices, and interfacial thermal resistance evaluations. Both methods rely on the thermoreflectance effect, where the surface reflectivity of a material undergoes slight changes with temperature. By optically probing transient or steady-state temperature responses, these techniques invert thermal parameters such as thermal conductivity, thermal diffusivity, and interfacial thermal conductance.

TDTR typically employs femtosecond or picosecond lasers as both the heat excitation source and the probe source. In experiments, a modulated pump beam periodically heats the sample surface, inducing a transient temperature rise. A probe beam irradiates the same area at varying time delays. Measuring the change in reflectivity yields the time-resolved temperature decay curve of the sample surface. To enhance the thermoreflectance signal, a metal transducer layer is usually deposited on the sample surface, with aluminum thin films being the most common choice. This metal layer serves as both the heat source for absorbing laser energy and the temperature-sensitive optical probe layer. Fitting the experimentally obtained transient signals with theoretical heat diffusion models allows for the simultaneous determination of multiple thermal properties of the sample and its interfaces. Figure 9(a) illustrates that the sample is typically coated with a metal thin film to act as a transducer. Figure 9(b) demonstrates the operating principle of TDTR, which measures the relationship between the time delay of pump and probe pulses arriving at the sample surface and the thermoreflectance response. In this process, the pump beam applies a periodic heat flux to the sample surface, while the probe beam detects the corresponding temperature changes through variations in reflectivity[67].

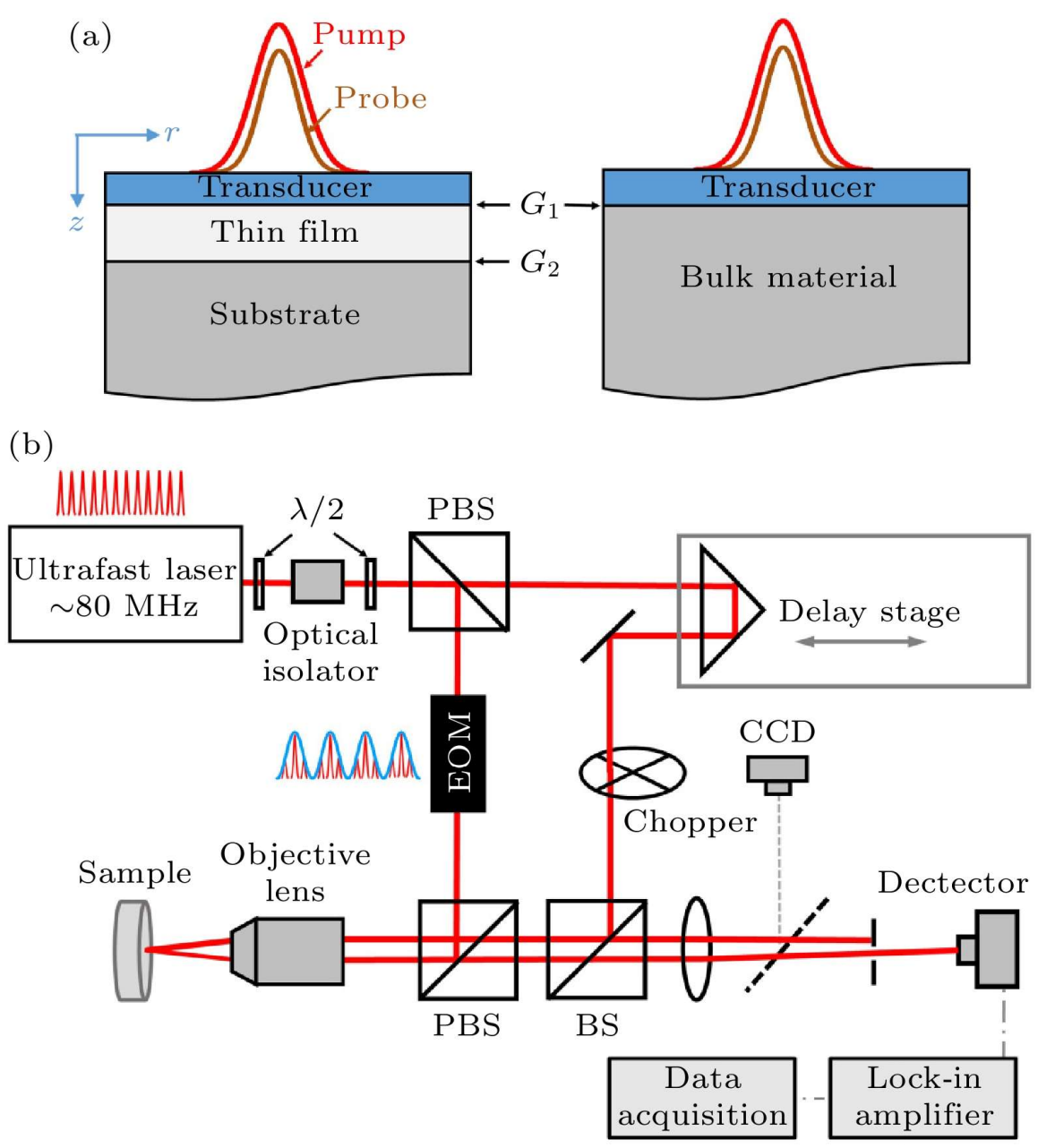

**Fig. 9 (a) Typical sample configuration for thin film and bulk material measured by TDTR; (b) schematic diagram of a typical transient thermal reflection device[67].**

Numerous scholars have made significant contributions to the extension of the TDTR method. For instance, Feser and Cahill[68] developed a TDTR scheme based on spatial offset or variable spot size. This approach enables independent measurement of anisotropic thermal conductivity in materials and expands the applicability of TDTR to complex heat transport systems. Wilson et al.[69] systematically measured the thermoreflectance and d$R$/d$T$ values of various metal transducer layers. Their work provides important guidance for selecting transducer materials and optimizing experimental design. Furthermore, Jiang et al.[70] proposed a dual-modulation frequency TDTR method. By combining high- and low-frequency signals, this method improves the accuracy of thermal conductivity measurements for thin films. It effectively enhances sensitivity and data reliability, particularly in extracting the thermal conductivity of high-thermal-conductivity thin films.

The primary advantage of TDTR lies in its high temporal resolution. It can directly probe heat transport processes on the picosecond to nanosecond scale. Therefore, it is particularly suitable for studying thin films, superlattices, and thermal boundary resistance (TBR). Additionally, TDTR can modulate the thermal penetration depth to some extent by varying the laser spot size, modulation frequency, or transducer layer thickness. This allows for selective characterization of heat transport behaviors at different scales. However, the experimental system for TDTR is relatively complex. It imposes strict requirements on optical path stability, the spatial overlap precision of pump and probe beams, and the mechanical stability of the time-delay line. These factors increase system complexity and debugging difficulty. Moreover, the data inversion process is sensitive to model assumptions, and strong correlations may exist between different parameters[71].

In contrast, FDTR uses a continuous-wave laser as the heat source. It applies sinusoidal modulation to the pump beam, generating periodic temperature oscillations on the sample surface. The probe beam synchronously measures the amplitude and phase information of reflectance changes as a function of modulation frequency. Schmidt et al.[72] systematically developed FDTR into a method for measuring thermal conductivity and heat capacity. Their work detailed how to obtain material thermal properties through frequency scanning and established the relationship between the thermal diffusion model and experimental response. Unlike TDTR, FDTR does not rely on time-delay scanning. Instead, it modulates the thermal diffusion length by changing the modulation frequency, thereby acquiring response characteristics of materials at different thermal penetration depths. Experimental data are typically presented as phase lag or amplitude attenuation versus frequency and are analyzed by fitting to a frequency-domain heat conduction model. A major advantage of FDTR is its relatively simple experimental implementation. It has lower requirements for laser pulse width and offers a better signal-to-noise ratio under high-frequency modulation conditions.

Furthermore, the frequency scanning approach gives FDTR certain advantages in distinguishing bulk thermal conductivity from interfacial thermal conductance. It is especially suitable for studying low-thermal-conductivity materials and multilayer structural systems. Regner et al.[73] further developed broadband FDTR (BB-FDTR) technology based on the FDTR platform. By extending the modulation frequency range, this technique enables the measurement of complex heat transport properties, such as the thermal conductivity accumulation function. This offers distinct advantages for investigating low-thermal-conductivity materials, TBR, and nanostructured systems. Yang et al.[74] proposed an extension of FDTR for microscopic imaging. This allows FDTR to simultaneously map different thermal parameters in multilayer structures. Their work demonstrates that FDTR is not only suitable for measuring the thermal conductivity of bulk materials and thin films but also effective in separating the contributions of interfacial thermal conductance and thickness. However, due to the lack of direct time-resolved information, FDTR has certain limitations in studying ultrafast thermal processes or strong non-equilibrium transport phenomena. Figure 10 shows a schematic diagram of FDTR systems using two different laser wavelengths.

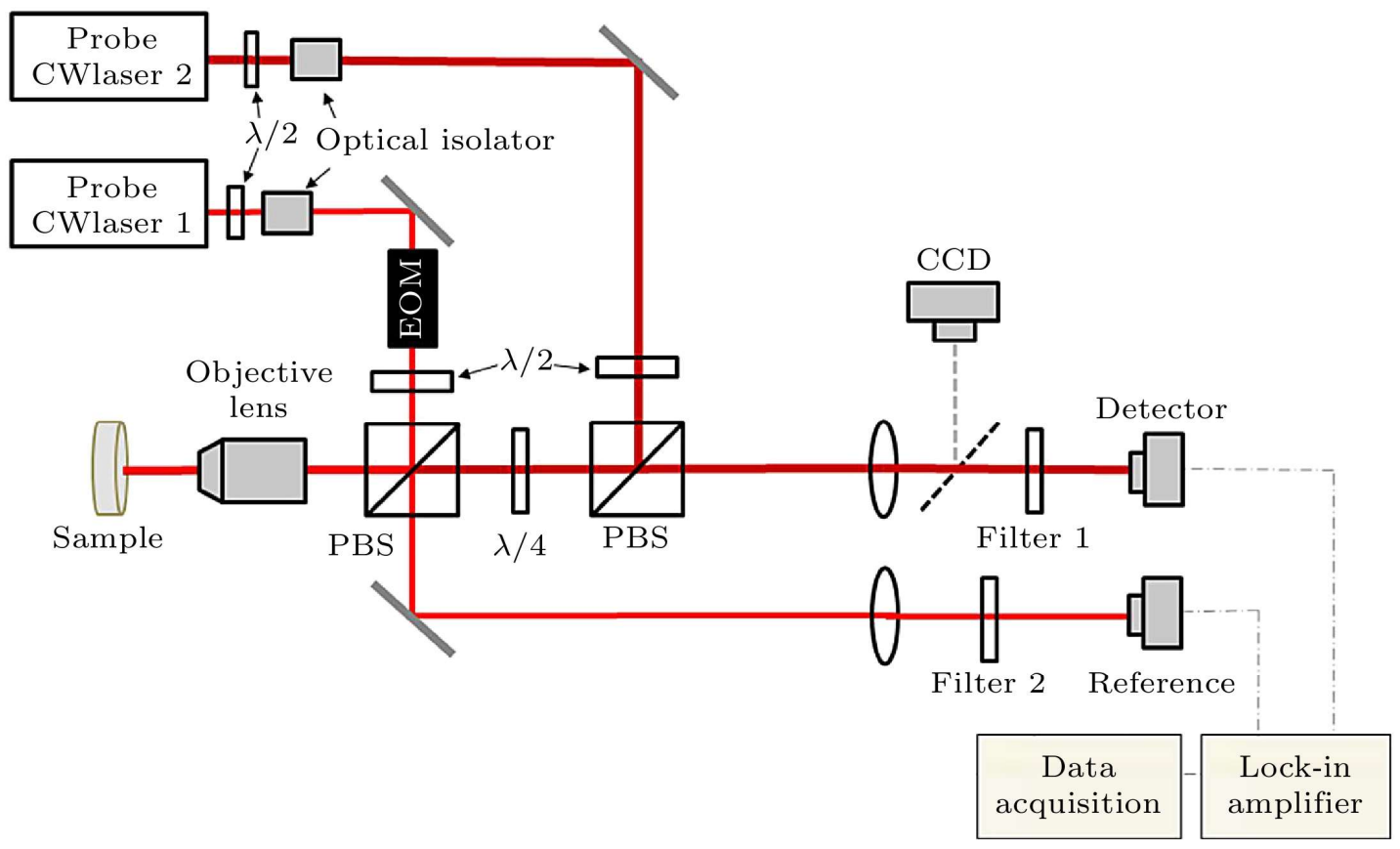

**Fig. 10 FDTR system based on two different wavelength continuous wave lasers[67].**

As shown in Figure 11, TDTR and FDTR exhibit distinct differences in experimental implementation and signal acquisition. However, both techniques rely on the thermoreflectance effect to probe the surface temperature response of materials[75]. TDTR directly captures the transient temperature decay of the sample in the time domain by varying the time delay between pump and probe pulses. This approach provides high temporal resolution, making it suitable for investigating ultrafast heat transport and thermal boundary resistance (TBR). In contrast, FDTR employs a continuous-wave laser with sinusoidal modulation of the pump beam. It acquires the amplitude and phase of the reflected signal by scanning the modulation frequency. The resulting measurements reflect the steady-state or quasi-steady-state thermal response of the sample in the frequency domain. This frequency modulation enables FDTR to effectively control the thermal penetration depth, offering advantages in distinguishing contributions from thin films and substrates. Overall, TDTR focuses on thermal dynamic processes in the time domain, whereas FDTR achieves spatial selectivity through frequency control. These two methods demonstrate strong complementarity in thermal property characterization.

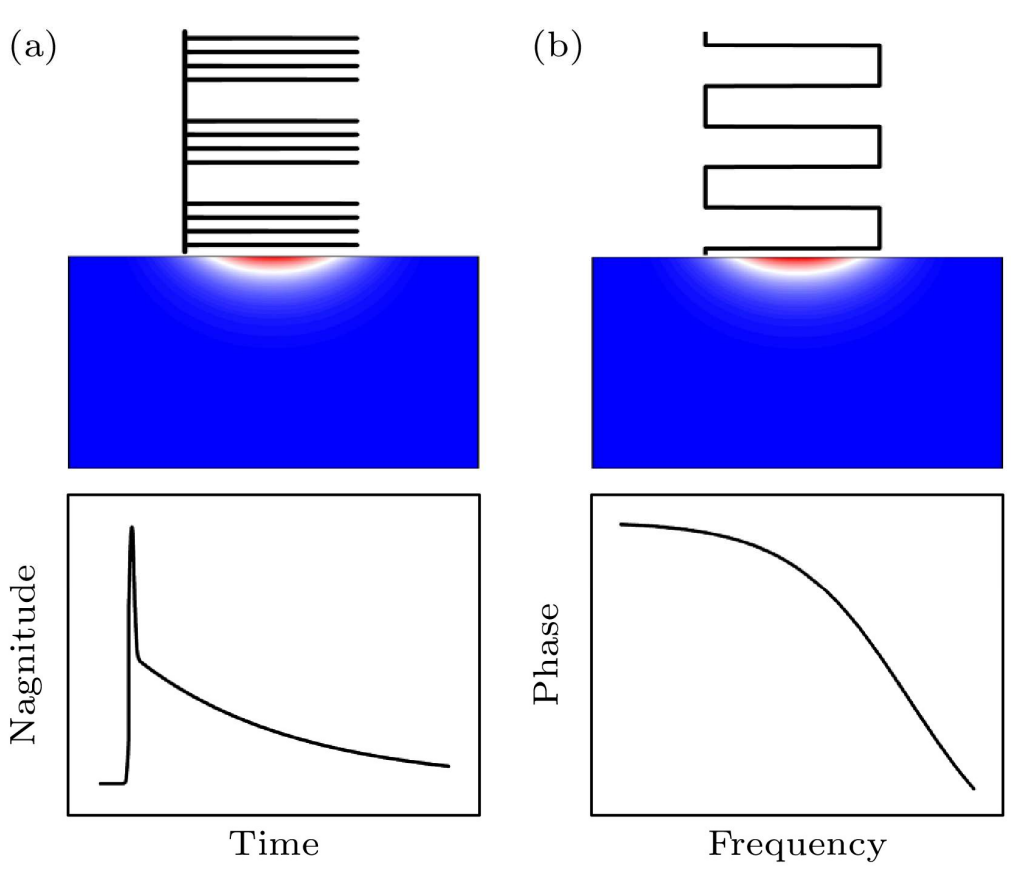

**Fig. 11 Characteristic excitations and corresponding responses for (a) TDTR and (b) FDTR[75].**

Overall, TDTR and FDTR share high consistency in physical principles but differ in experimental implementation and data analysis strategies. TDTR emphasizes time resolution, making it suitable for ultrafast and high-precision temporal measurements. In contrast, FDTR achieves thermal scale selection through frequency modulation, demonstrating excellent stability and sensitivity under steady-state or quasi-steady-state conditions. In practical research, these two methods are often used complementarily to enhance the reliability and accuracy of thermal property extraction. With continuous advancements in experimental techniques and theoretical models, TDTR and FDTR have become indispensable tools for studying heat transport at the micro- and nanoscales.

Beyond TDTR and FDTR, which are based on the optical pump-probe principle, spectroscopic characterization methods using advanced electron microscopy have recently shown unique advantages in nanoscale heat transport research. Scanning transmission electron microscopy combined with electron energy loss spectroscopy (STEM-EELS) can probe phonon excitations and local energy distributions with sub-nanometer spatial resolution. This capability enables direct characterization of temperature fields and interfacial heat transport behavior near heterogeneous material interfaces. This method overcomes the spatial resolution limits of traditional optical measurements. It provides crucial experimental evidence for revealing phonon-dominated energy transfer mechanisms at nanostructured interfaces, serving as a powerful complement to macroscopic averaging techniques such as TDTR[76].

### 3.2.3 Microchannel-Cooling Experimental Platforms

For chip cooling structures employing liquid cooling, the core objective of the microchannel cooling experimental platform is to quantitatively characterize flow and heat transfer properties. Key parameters include channel pressure drop, mass flow rate, and local and average heat transfer coefficients. The channel pressure drop and flow characteristics reflect microscale flow resistance and stability, forming the basis for evaluating cooling system energy consumption and reliability. The heat transfer coefficient and its variation with heat flux directly determine the heat dissipation capacity of microchannel cooling schemes under high heat flux conditions. To achieve precise measurement of these parameters, the experimental system typically integrates microchannel cold plates fabricated using micro-electromechanical systems (MEMS) technology. These cold plates allow for high controllability in geometric dimensions, channel morphology, and surface conditions, providing a reliable platform for systematic research. During experimental testing, high-precision pressure sensors and mass flow meters monitor flow parameters in real time. High-precision temperature sensors, such as thin-film resistance thermometers or thermocouple arrays, are arranged on the chip simulated heat source or cold plate surface to obtain temperature distributions and their dynamic changes. Additionally, high-speed camera systems are often introduced to visualize two-phase flow patterns and boiling behaviors within microchannels, including processes such as bubble nucleation, growth, coalescence, and detachment. This provides intuitive evidence for elucidating microscale boiling heat transfer mechanisms. By comparing experimental results for pressure drop, heat transfer performance, and flow pattern evolution with computational fluid dynamics (CFD) numerical simulations, researchers can effectively validate physical assumptions regarding two-phase flow and boiling heat transfer in the models. This comparison also improves the accuracy and reliability of critical heat flux predictions[77].

Wu et al.[78] recently proposed an embedded, jet-enhanced manifold microchannel structure for cooling chips at high heat flux. Their composite microfluidic architecture comprises a manifold layer, an intermediate microjet layer, and a microchannel layer with zigzag sidewalls. It removed heat fluxes of up to 3000 W/cm$^2$ while maintaining a low pumping-power density. Combining microchannels, manifolds, and jets reduces the overall pressure drop and increases local heat-transfer coefficients, providing an experimental platform for the thermal management of high-power electronic, power-electronic, and radio-frequency devices. Compared with conventional single-channel or simple microchannel structures, it offers better thermal-hydraulic performance and energy efficiency. Table 4 compares methods for measuring thermophysical properties and thermal boundary resistance.

**Table 4** Comparison of characteristics of thermal properties and TBR measurement experimental methods in chip thermal management.

| Measurement Scale | Experimental Method | Main Contribution | Reference |
|---|---|---|---|
| Thin film/micrometer scale | 3ω method | Thin-film thermal conductivity and TBR, applicable to thin films and microelectronic structures | [63] |
| Nano-micrometer scale | TDTR | TBR and thermal diffusivity, applicable to interfaces of various materials | [67] |
| Nano-micrometer scale | FDTR | TBR and thermal diffusivity exhibit high parameter sensitivity, and their phase signals are sensitive to experimental noise, signal-to-noise ratio, and lock-in amplifier stability, requiring control through frequency optimization and uncertainty analysis | [75] |
| Chip level | Microchannel cooling experimental platform | Pressure drop, flow rate, and heat transfer coefficient, enabling direct validation of liquid cooling and two-phase flow models | [77] |

### 3.3 Joint Simulation-Experiment Validation

Experimental measurements provide data for model calibration and error analysis. Studies of structures such as GaN HEMTs, HBM-3D stacks, and 3D NAND show that comparison of TR or Raman measurements with FEM temperature fields can reveal errors in the power map, uncertainty in TBR, and parasitic heat-flow paths. In HBM structures, for example, measured hotspot temperatures are often 10%-15% higher than simulated values, primarily because the actual thermal-interface-material (TIM) thickness and material properties differ from the modeled values. Joint validation improves overall predictive accuracy and supports optimization of the cooling design.

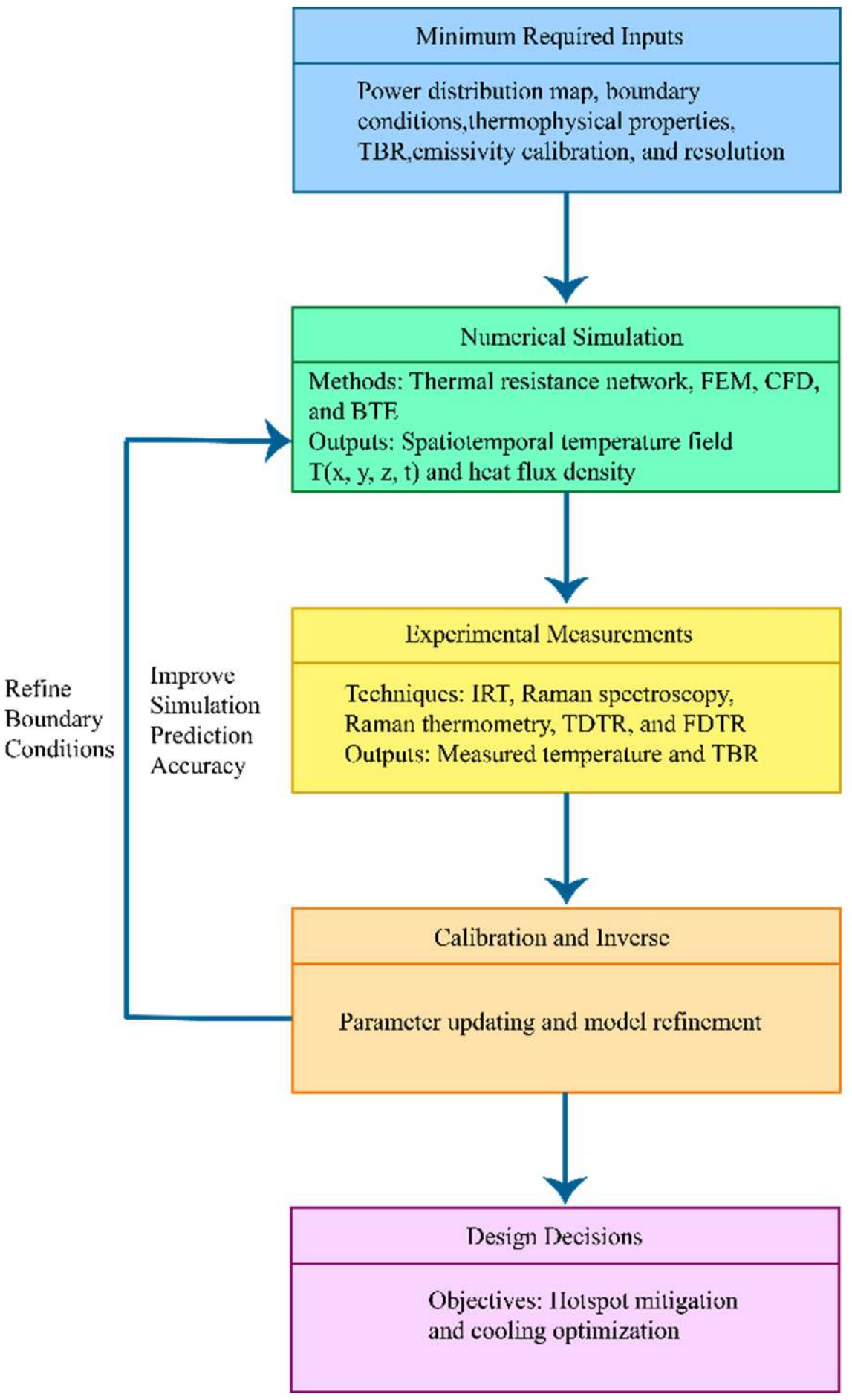

**Fig. 12 Simulation-measurement-calibration closed-loop flowchart.**

# 4 Technical Bottlenecks

Although numerical simulation and experimental measurement have advanced substantially, chip thermal design still faces several bottlenecks as power densities rise, packages become more complex, and application requirements change rapidly. At the same time, simulation, measurement, and model calibration are increasingly being integrated into a closed loop, as shown in Figure 12. The following discussion considers three categories of limitations: simulation, experiment, and system-level cooling.

## 4.1 Simulation Bottlenecks

From a simulation perspective, the coupling of macroscopic and microscopic heat conduction, the solution of transient high heat flux density fields, and the uncertainty of power maps and material parameters are key factors limiting accurate prediction. The challenges mainly fall into three aspects. First, when chip packaging adopts 3D stacking, chiplet, or embedded cooling structures, full-chip transient thermal simulations often require hundreds of millions of grid cells. Combined with thermal-fluid-structural coupling models, the solution time can extend from hours to days or even weeks. This not only prolongs the design cycle but also makes "rapid iteration" nearly impossible[79]. Second, as shown in Figure 13, significant non-Fourier effects occur when the characteristic scale approaches the phonon mean free path. The figure summarizes the heat transfer mechanisms and applicable scale ranges. The scale gap between macroscopic simulations and microscopic models remains difficult to bridge effectively. In actual chips, phonon-level non-Fourier heat transfer and thermal boundary resistance (TBR) effects couple with large-scale thermal diffusion processes at the package level. Traditional models struggle to account for both, leading to significant deviations in temperature field predictions. Existing reviews indicate that although simulation libraries are continuously expanding, local prediction errors in real dynamic load environments remain between 10% and 20%[1,80]. Third, simulation accuracy highly depends on parameters such as power maps, material thermal properties, and TBR. However, these parameters often exhibit significant variability in actual packaging. For instance, thermal interface materials degrade in performance due to service life, temperature cycling, and mechanical stress. This constrains the credibility of simulation outputs. Industry reports indicate that thermal interface materials have become a non-negligible bottleneck in the "thermal path." As shown in Figures 14 and 15, errors in power inversion are more pronounced in the presence of thermal noise. Poorer parameter selection exacerbates these errors. The uncertainty of power maps and interface parameters is amplified by thermal issues, making inversion or prediction results highly sensitive to noise. Thus, at the simulation technology level, achieving the goal of "predicting the full-chip transient temperature field within hours" still requires overcoming the three challenges mentioned above.

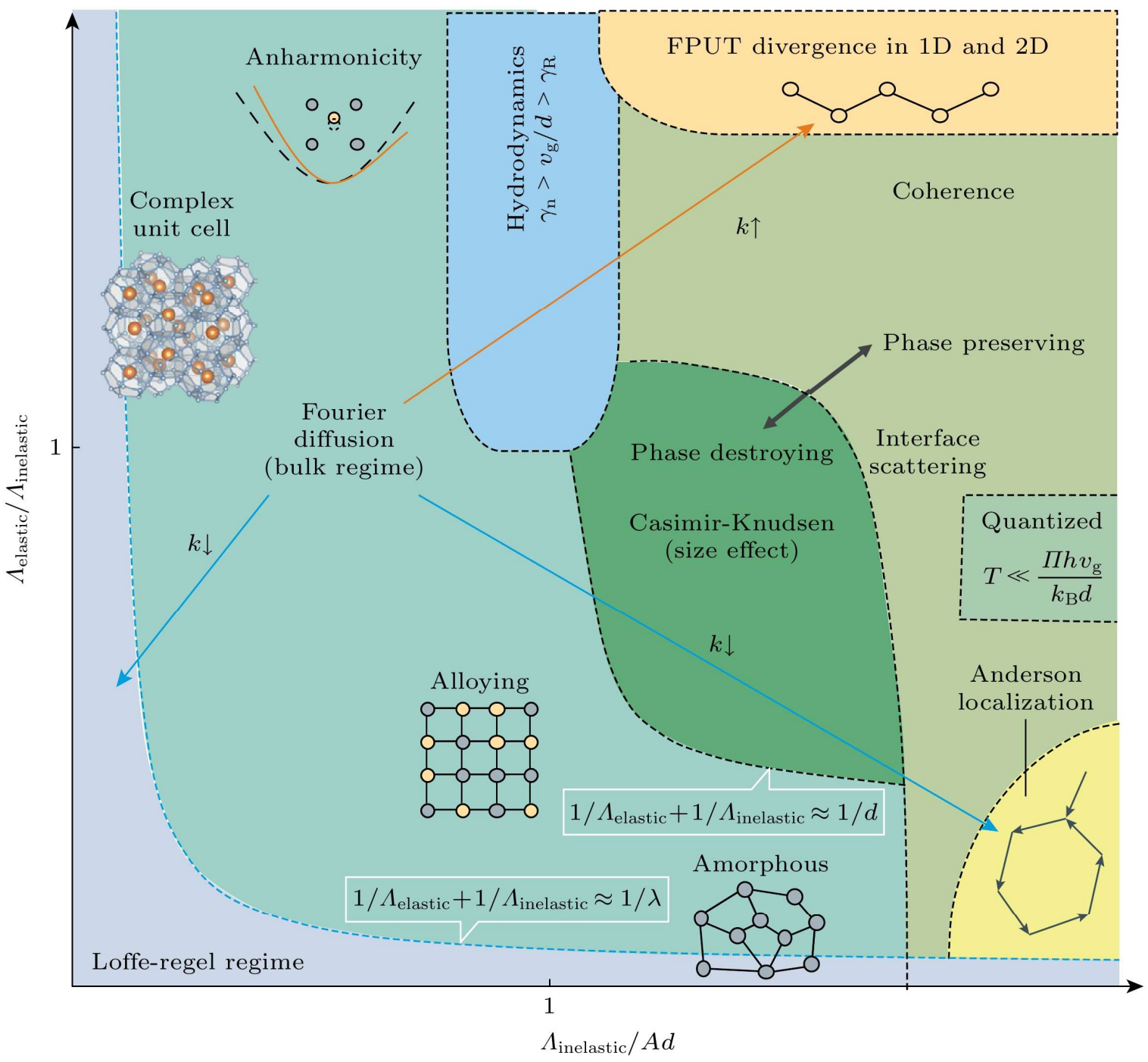

**Fig. 13 Overview of heat conduction mode[81].**

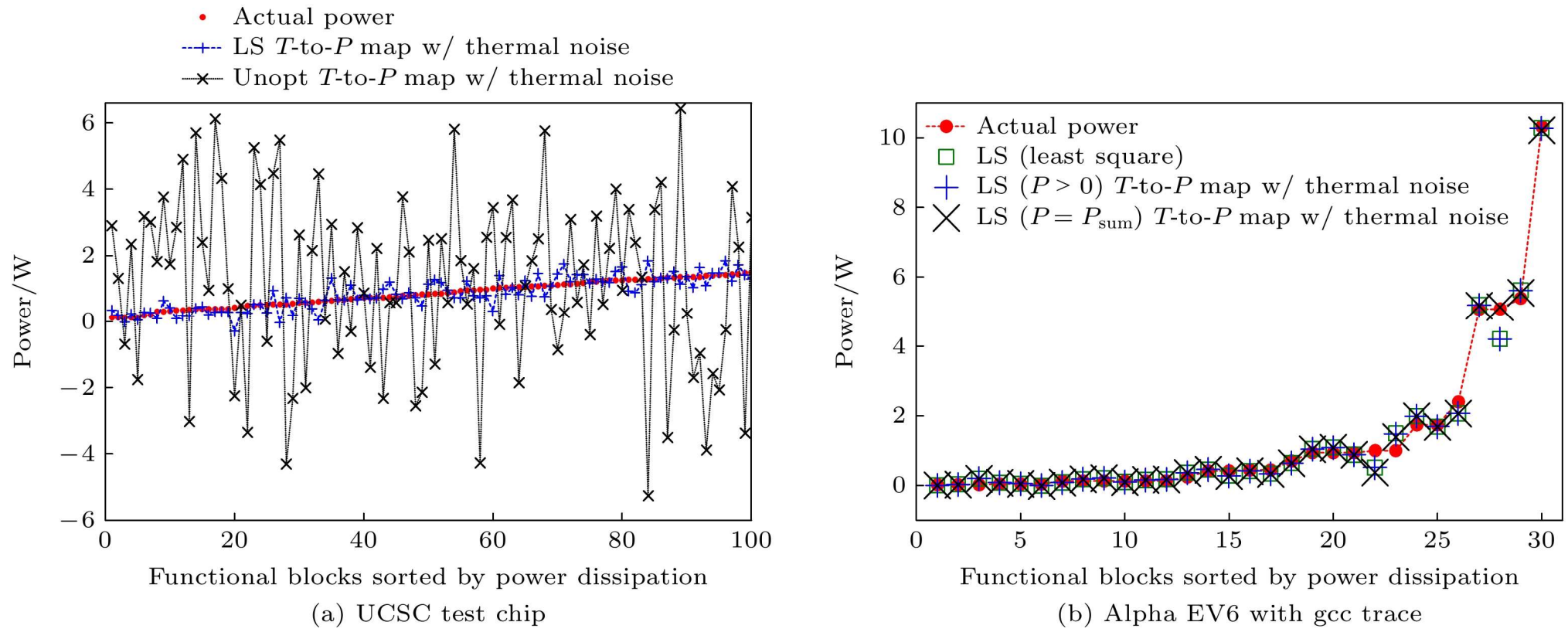

**Fig. 14 Steady-state inversion mapping[82].**

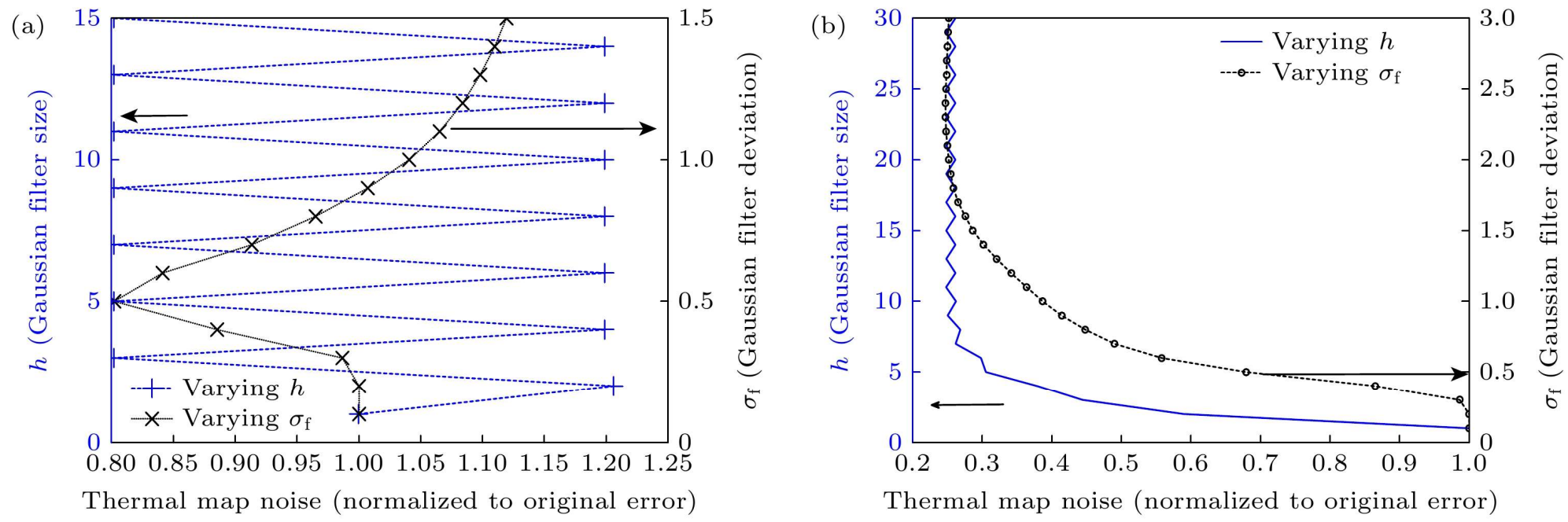

**Fig. 15 The influence of Gaussian filter parameters under different noise conditions[82]: (a) Thermal noise, $\mu$ = 0, $\sigma$ = 5%; (b) thermal noise, $\mu$ = 0, $\sigma$ = 25%**

## 4.2 Experimental Bottlenecks

From the perspective of experimental measurement, multiple constraints exist. It is difficult to balance spatial and temporal resolution. Experimental reliability is poor under high heat flux density conditions. Issues regarding the cost and repeatability of thermal boundary resistance (TBR) measurements are prominent. For scenarios where the hotspot scale is less than 1 μm and the temperature transition time is less than 1 μs, current techniques such as infrared thermography, thermoreflectance, and Raman spectroscopy struggle to achieve extreme spatial and temporal resolution simultaneously in a single experiment. As shown in Figure 16, the measurable length scales and achievable temperatures of different thermal measurement methods are compared. Thermoreflectance performs well in measuring micro- and nanoscale properties, such as TBR. Its measurable temperature range is primarily influenced by factors such as transducer material stability, oxidation, and interfacial diffusion, rather than limitations of the technical principle itself. Even when using aluminum, which has a relatively low melting point, as the transducer, thermoreflectance measurements can operate stably up to approximately 300 °C under appropriate experimental conditions. This temperature range covers the practical application requirements of most chips and microelectronic devices. Infrared imaging is limited by wavelength, resulting in a resolution often at the scale of several micrometers. Although Raman methods can achieve micrometer-level resolution, their slow scanning speed makes them unsuitable for large-area real-time monitoring. Current temperature measurement techniques still exhibit certain limitations when handling dynamic loads and rapid hotspot changes[80]. When power density exceeds the kilowatt per square centimeter level, such as in 3D-stacked high-bandwidth memory scenarios, experimental samples are easily destroyed due to thermal runaway. Fluid cooling and microchannel boiling experiments also face issues such as liquid-gas interface instability and uneven coolant distribution. These problems affect measurement repeatability and reliability[83]. Key thermophysical parameters, such as thin-film thermal conductivity and TBR, are typically obtained using techniques like 3$\omega$ and time-domain thermoreflectance (TDTR). Although the associated experimental equipment is costly and imposes strict requirements on experimental conditions and operational standards, these methods demonstrate good repeatability and reliability under reasonable experimental design and uncertainty control. They have been widely used for the calibration and validation of heat transport models[84]. Therefore, in terms of experimental platform construction and the improvement of measurement systems, there is an urgent need for technological breakthroughs that offer higher resolution, higher reliability, and controllable costs.

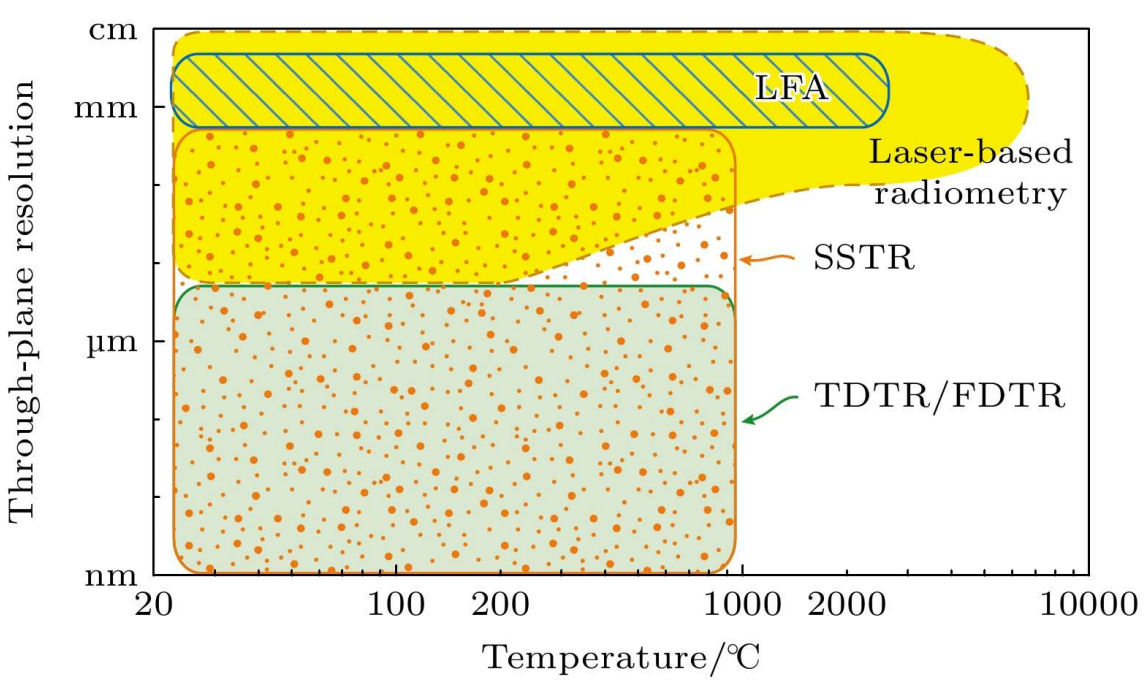

**Fig. 16 State diagram of non-contact thermal transport measurement[85].**

## 4.3 System-Level Bottlenecks

From a system-level perspective, traditional air-cooling and heat pipe systems have approached their physical limits, while the engineering implementation of novel cooling solutions remains challenging. As heat flux densities exceed kilowatts per square centimeter, dissipation pathways relying on air or heat pipes become increasingly ineffective. Although microchannel liquid cooling and immersion cooling offer more efficient thermal channels, they have not yet been widely adopted for mass production due to challenges in packaging, reliability, maintenance, and cost[86].

Selection of a system-level cooling solution depends not only on heat-transfer capacity but also on engineering constraints such as pressure drop, pumping power, long-term reliability, and manufacturability. More than 20% of high-performance-computing systems have begun to deploy direct-to-chip liquid cooling or immersion cooling in response to rising power densities. These systems, however, add pumping power and maintenance complexity, requiring a trade-off between temperature control and energy efficiency. Under experimental conditions, embedded microchannels have removed heat fluxes of approximately 3000 W/cm² at a pumping-power density of about 0.9 W/cm²[78]. Pressure drop, clogging risk, and integration with package fabrication nevertheless remain major barriers to large-scale adoption[87]. Chip thermal simulations should therefore represent cooling solutions explicitly in the boundary conditions and include pumping power, reliability, and package compatibility as system-level optimization constraints.

As shown in Figure 17, embedded microchannel cooling integrates microscale liquid cooling channels directly into the backside of the chip package or the interposer. This approach ensures close contact between the coolant and hotspots, significantly shortening the thermal path and reducing thermal resistance. This integrated structure serves as a key support for the future design framework of "cooling-packaging-chip co-optimization," offering the potential to overcome the physical limits of traditional air-cooling and heat pipe solutions. Thermal management is no longer an optional add-on in the post-design phase; instead, it should be considered in parallel with chip architecture, packaging, and system cooling integration. However, most current design processes still treat thermal design as a subsequent correction step, resulting in passive thermal paths and reduced optimization margins. Industry commentators have stated that "thermal management has become a decisive factor in advanced packaging." In summary, system-level bottlenecks necessitate a transition from "feasibility of cooling technology" to a new stage of "cooling-packaging-chip co-optimization."

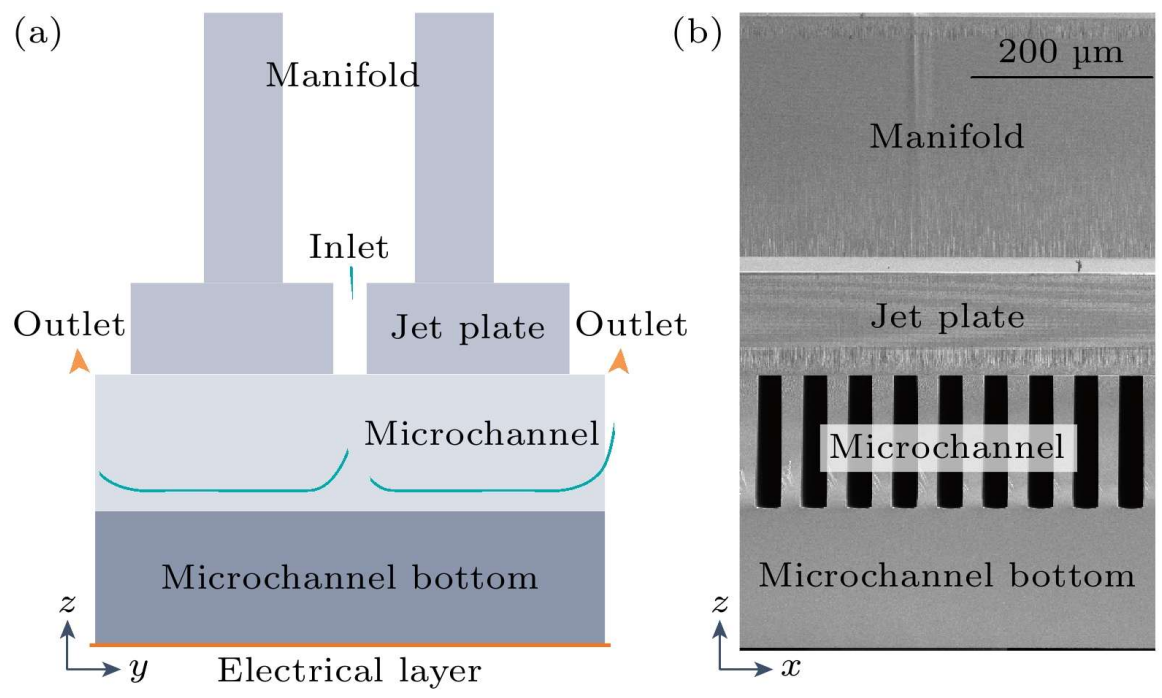

**Fig. 17 3D embedded microchannel diagram[88].**

# 5 Future Research Directions

As chip power density continues to rise and packaging evolves rapidly toward 3D

heterogeneous integration, traditional thermal management frameworks struggle to meet the stringent demands of future high-performance computing, artificial intelligence, and 6G communication scenarios. To overcome the triple bottlenecks in simulation, experimentation, and systems described in Section 4, future research must advance synergistically across multiple dimensions, including theoretical modeling, thermal materials, cooling structures, and design paradigm innovations.

## 5.1 Emerging Simulation Technologies

Future thermal simulation is expected to move from conventional partial-differential-equation-based 'solver' workflows toward data-driven 'predictor' workflows. Deep-learning surrogates can be trained on high-fidelity FEM, CFD, or Boltzmann-transport data to generate temperature fields rapidly for a wide range of materials and geometries. Figure 18 illustrates a hierarchical graph-neural-network framework for multiscale thermal modeling. Studies have shown that U-Net and graph-neural-network architectures can achieve speedups of $10^3$-$10^4$ while retaining approximately 95% accuracy[88]. Such end-to-end thermal analysis could enable real-time design assessment, dynamic temperature control, and online optimization. At device dimensions of tens of nanometers, quantum effects such as phonon coherence and nonlocal heat conduction also become increasingly important. BTE models based on the relaxation-time approximation cannot fully describe phonon interference or quantum boundary scattering. Developing quantum heat-transport frameworks coupled to first-principles methods such as density functional theory (DFT) and density functional perturbation theory (DFPT) is therefore an important direction. The phonon-spectrum modeling approach proposed by Chen[89], for example, is being extended to predict coherent heat transport in superlattices and two-dimensional interfaces. Figure 19 presents experimental observations of coherent phonon transport. Such high-fidelity models will provide a physical basis for optimizing advanced package structures.

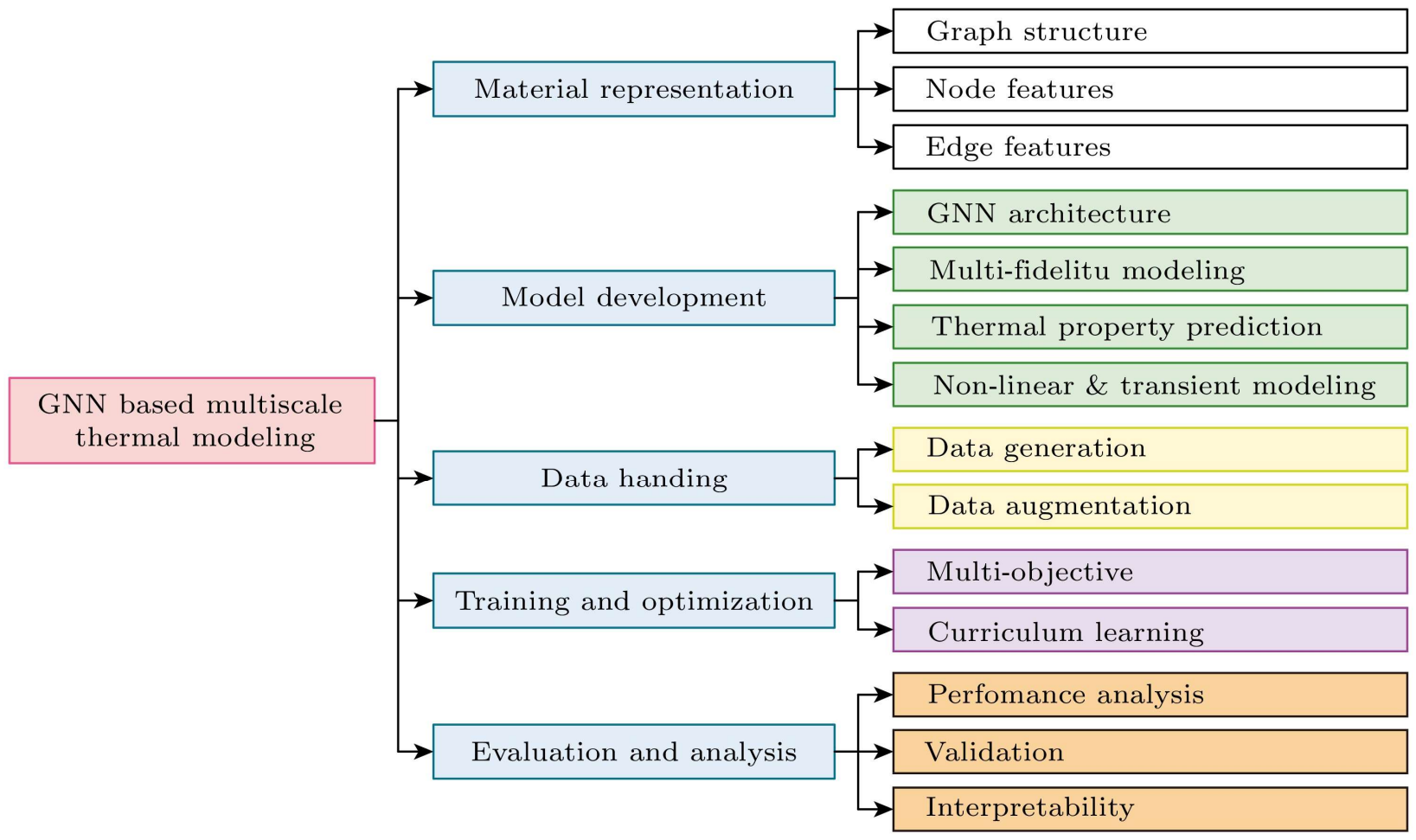

**Fig. 18 Hierarchical structure[90].**

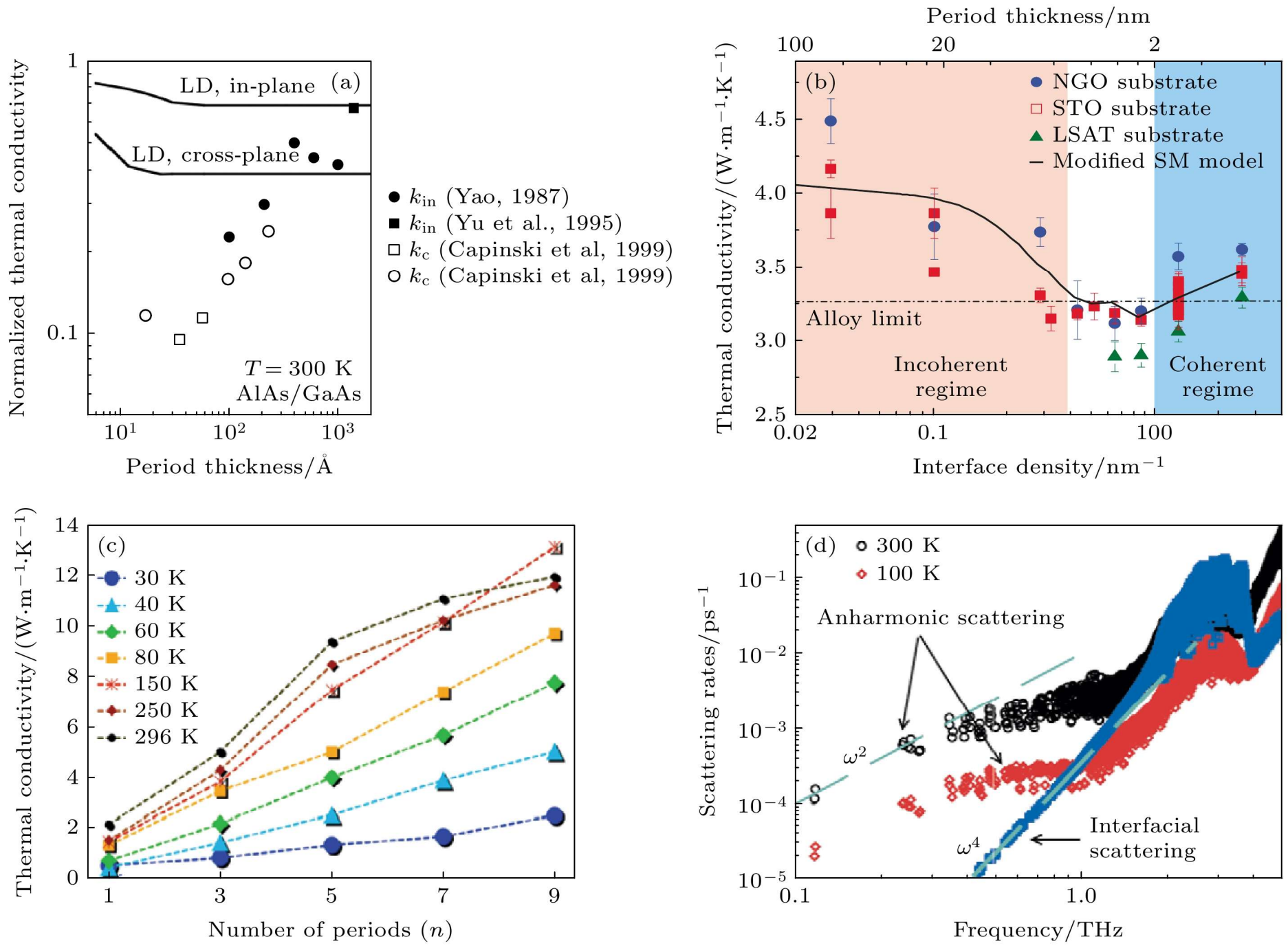

**Fig. 19 Coherent phonon transport: (a) The process of calculating the ideal lattice thermal conductivity through lattice dynamics simulation (group velocity, confinement and tunnelling effects included) in both the cross- plane and in- plane directions (solid lines)[91]; these simulations do not agree with experimental data (squares and circles)[92–94]; (b) the thermal conductivity of the $SrTiO_3/CaTiO_3$ oxide superlattice varying with the interface density; (c) the thermal conductivity of the GaAs/AlAs superlattice with a fixed periodic thickness and an increase in the number of single periods; (d) the first-principles phonon scattering simulation in the GaAs/AlAs superlattice[81].**

## 5.2 Advanced Cooling Technologies

Cooling is shifting from conventional heat removal at the chip surface toward active cooling within the chip. Embedded microchannel liquid cooling forms channels directly in the silicon wafer, greatly shortening the heat-transfer path between the coolant and heat-generating regions. Experiments generally use deionized water or dielectric coolants and conduct steady-state or transient tests on silicon test chips with integrated heaters under controlled flow rates and pressure drops. With suitable channel geometries and pumping inputs, heat-removal capacities above 1 kW/cm² have been demonstrated in the laboratory, and repeated tests have confirmed the reproducibility of the reported pressure drops, pumping requirements, and thermal performance[62]. Two-phase cooling, including microjet evaporation, can provide very high heat-transfer coefficients because of the large latent heat of vaporization and is considered a key option for future high-heat-flux devices[95]. Co-design of chips and liquid-cooling components, long-term sealing, and reliability nevertheless require further study. Advanced high-thermal-conductivity materials, including graphene, hexagonal boron nitride, MXenes, and their composites, are also extending heat-spreading capabilities.

Hexagonal boron nitride, for example, can exhibit in-plane thermal conductivity above 400 W/(m·K), making it a promising candidate for thermal interface materials, heat spreaders, and interface fillers[96,97].

Graphene and boron nitride have been studied in detail. The structure and simulation results in Figure 20 illustrate how graphene can enhance heat spreading. To clarify the anisotropy of phonon transport across graphene/boron-nitride heterointerfaces, prior work analyzed the phonon power spectrum of a 75 nm-long system and calculated partial phonon densities of states (PDOS). Figure 21 shows PDOS curves for carbon and boron-nitride atoms under forward and reverse heat flow. The in-plane vibrational modes at different atomic positions remain nearly unchanged when the heat-flow direction is reversed, indicating only a limited in-plane contribution to directional transport. By contrast, out-of-plane modes near the interface differ markedly between the two directions, demonstrating the importance of interfacial phonon coupling and mode mismatch. The direction-dependent heat transport at these heterointerfaces therefore arises from mismatched vibrational modes and highlights the role of interfacial phonon coupling in controlling thermal conduction[98]. Future materials research should emphasize interface engineering, orientation control, and defect control to improve package integration and long-term stability.

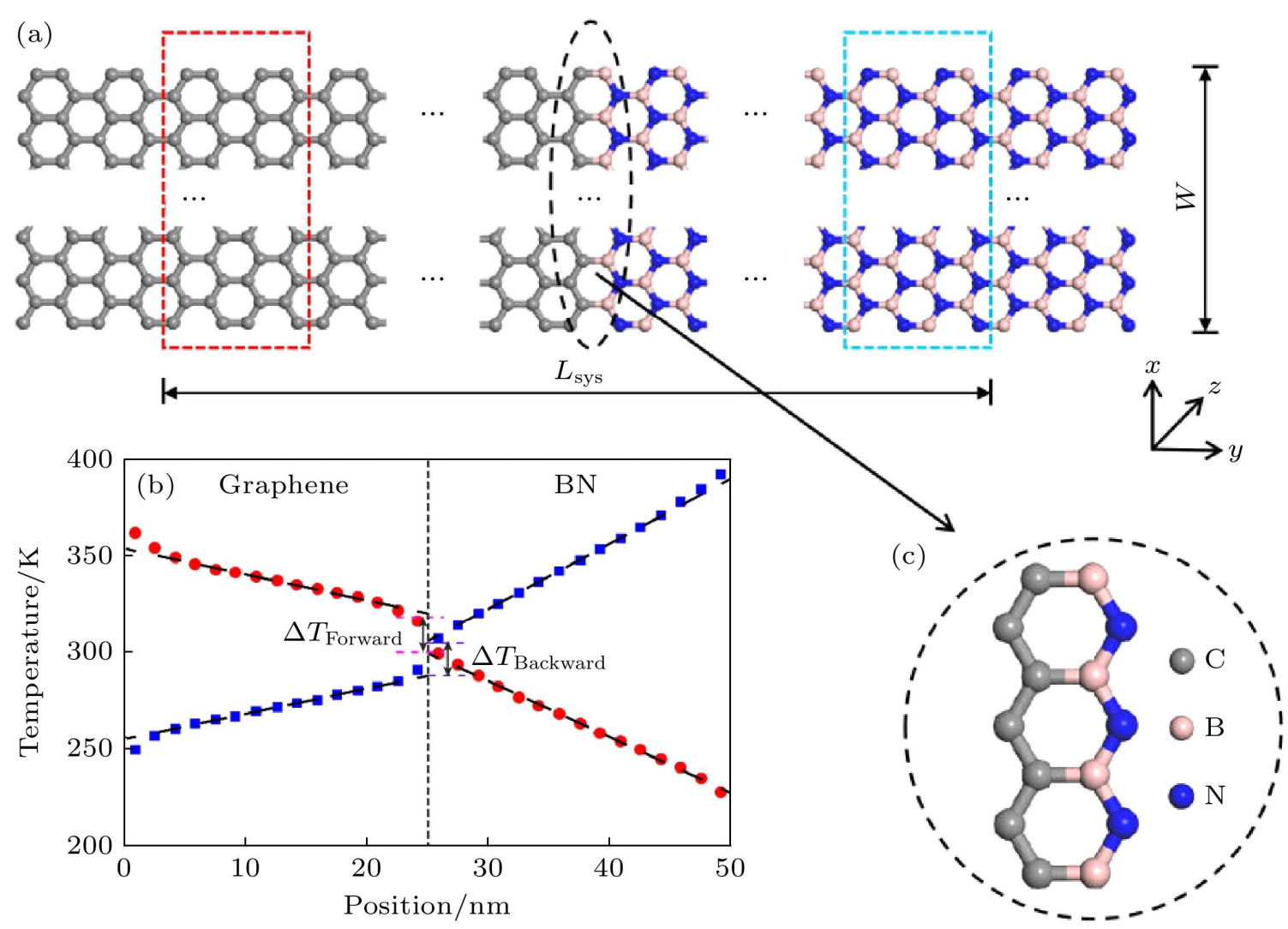

**Fig. 20 (a) Nanosheet thermal diode simulation system, it is formed by connecting the graphene and boron nitride (BN) in equal proportion, red and blue dashed lines represent heating area and cooling area, respectively, heat flows from graphene to BN (forward) or from BN to graphene (backward); (b) steady-state temperature profile of the 50 nm-long system, the red data points represent the forwards direction, while the blue ones represent the backwards direction; (c) interface of C—B—N bond formation[98].**

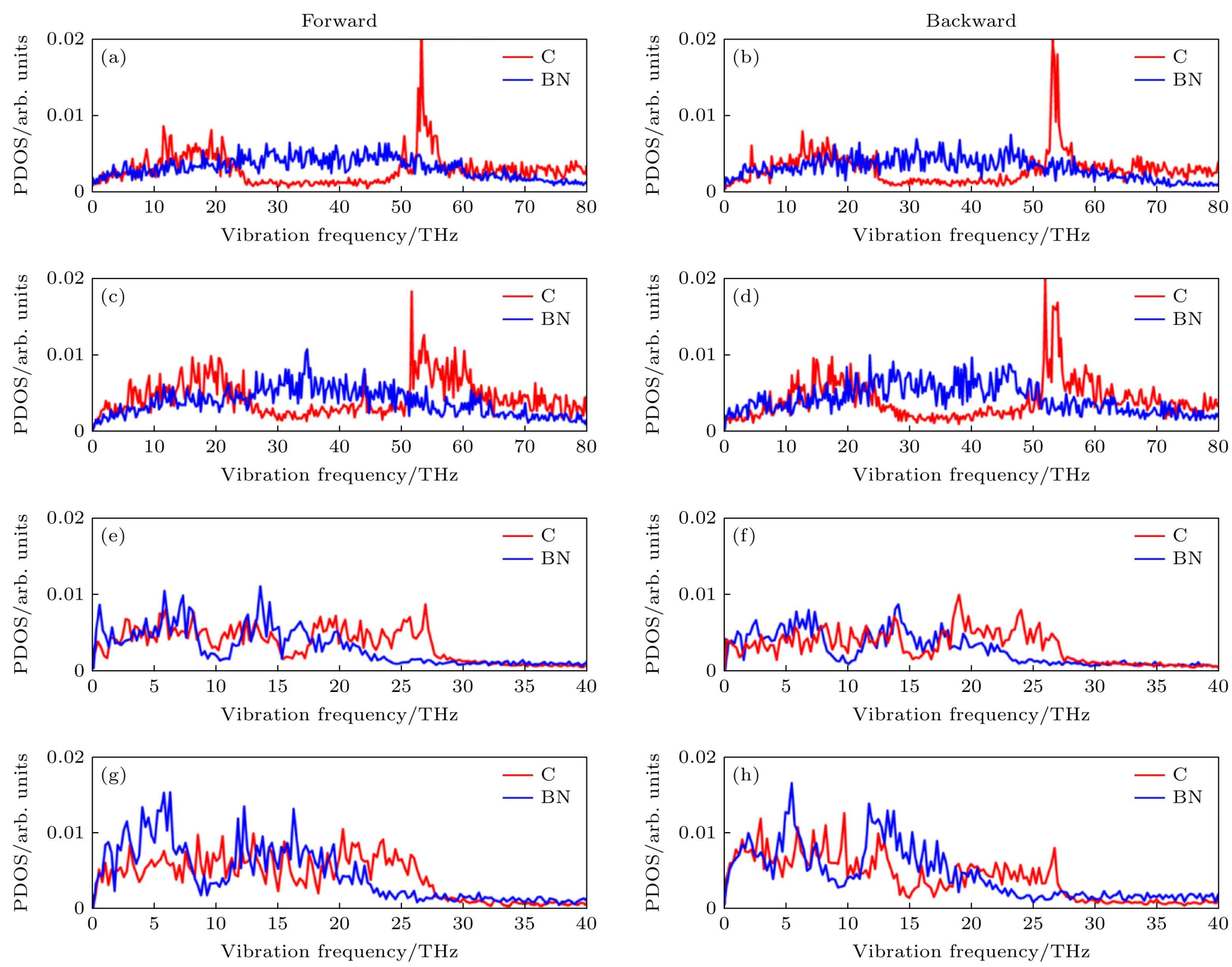

**Fig. 21 PDOS curves of carbon and boron nitride atoms in the forward and backward directions[98]: (a), (b) In-plane direction of the middle region; (c), (d) in-plane direction at the interface; (e), (f) out-of-plane direction of the middle region; (g), (h) out-of-plane direction at the interface.**

From an engineering perspective, this study highlights the potential value of interfacial phonon engineering in thermal management of two-dimensional materials. Although thermal rectification itself is not a direct objective of chip heat dissipation, the modulation of out-of-plane phonon modes at interfaces provides a physical basis for reducing interfacial thermal resistance and optimizing anisotropic thermal diffusion paths. These findings offer insights for structural design of graphene and related two-dimensional materials in chip-level heat dissipation and packaging applications.

## 5.3 Multiphysics Co-Optimization

Thermal design has traditionally been deferred to late-stage correction, creating a disconnect between chip architecture and cooling. Future workflows should integrate thermal management into early design stages and apply thermal-aware co-optimization from transistors, circuits, and physical layout through package cooling. Optimizing the placement of temperature-sensitive regions, reducing hotspot coupling, and adjusting TSV density and location can reduce overall chip thermal resistance. Co-optimizing electrical, thermal, and mechanical interactions can also improve the reliability of interconnects and solder joints. Studies indicate that electrothermal co-optimization can extend chip lifetime by 20%-30% while mitigating the frequency loss caused by dynamic thermal-management strategies[41,99].

## 5.4 High-Spatiotemporal-Resolution Measurements and On-Chip Sensor Networks

In terms of experimental characterization, future measurement technologies will continue to develop towards "high spatial resolution, integrability, and low power consumption." On one hand, ultrafast pump-probe techniques based on femtosecond lasers, combined with advanced spectroscopic methods such as synchrotron radiation, are expected to improve time resolution to the picosecond or even sub-picosecond level and advance spatial resolution to the hundred-nanometer scale. This will enable direct observation of transient heat transport processes within devices. On the other hand, high-spatial-resolution experimental techniques based on electron microscopy have demonstrated unique advantages in recent years. For example, phonon spectrum and energy distribution measurement methods combining Scanning Transmission Electron Microscopy-Electron Energy Loss Spectroscopy (STEM-EELS) can characterize local heat transport behavior near interfaces at the nanometer or even sub-nanometer scale. These methods provide new experimental pathways for understanding interfacial thermal resistance and non-uniform thermal diffusion in complex heterogeneous structures.

Advances in micro- and nanofabrication and sensing are also enabling integrable on-chip sensor arrays and distributed temperature-sensing networks for in situ, real-time monitoring during device operation. These measurements can validate multiscale heat-transport models and continuously update AI-based thermal-prediction and optimization algorithms. They therefore support closed-loop "measurement-simulation-control" thermal management and offer new options for reliability design in high-power-density chips and advanced packages.

## 5.5 Boiling-Based Cooling Technologies

Boiling-based cooling, which exploits phase-change heat transfer, is considered a promising option for future ultra-high-power chips because of its very high heat-transfer coefficients and its potential to accommodate extreme heat fluxes[100-103]. Compared with single-phase liquid cooling or air cooling, absorption of latent heat during liquid-to-vapor phase change can substantially reduce junction temperature and increase heat-removal capacity per unit area by an order of magnitude. Boiling-based cooling has therefore attracted growing interest for high-performance computing chips, power-electronic devices, and 3D integrated packages.

At chip scale, extensive research has focused on surface engineering and boiling-enhancement mechanisms. Micro- and nanostructures, porous coatings, and biomimetic wettability gradients on heated surfaces can control bubble nucleation, promote liquid replenishment, delay dryout, increase critical heat flux, and reduce interfacial thermal resistance[101,102,104]. Combining boiling with microchannel structures is also promising for highly integrated packages because it can balance high heat-removal capacity with system compactness in a limited volume.

On the other hand, as advanced packaging technologies evolve towards chiplets and 3D stacking, the dense distribution of local hotspots, increased number of interface layers, and significantly enhanced multiphysics coupling effects among heat, fluid, mechanics, and electricity pose numerous challenges to the practical application of boiling heat dissipation[105,106]. For instance, there is currently no unified theoretical description for the evolution laws of bubble dynamics in microscale confined spaces. The impact of interfacial thermal resistance on boiling stability in multi-layer packaging structures remains to be quantitatively assessed. Additionally, the reliability, contamination, and degradation of surface microstructures under long-term operating conditions have become important factors restricting engineering applications[107,108].

Future development of boiling-based cooling requires close integration with multiscale modeling and intelligent optimization[104,109]. Fundamental studies of micro- and nanoscale bubble nucleation, interfacial phase-change heat transfer, and flow instability should be combined with high-resolution experiments to establish models suitable for device-level prediction[110]. Machine-learning-assisted inverse design of surface structures, topology optimization of cooling channels, and optimization of operating conditions may move the field beyond experience-driven design[111,112]. Joint consideration of boiling, advanced thermal interface materials, embedded cooling structures, and system-level thermal-management architectures will be an important trend for next-generation high-power chips.

# 6 Conclusion

As integrated circuits evolve toward higher power density, 3D packaging, and heterogeneous integration, chip thermal design has become a central constraint on performance, reliability, and lifetime. This article has reviewed the principal numerical simulation methods and experimental measurement techniques used in chip thermal management, analyzed multiscale heat transport from packages to nanoscale interfaces, and summarized the main challenges associated with computational scale, multiphysics coupling, thermal-boundary-resistance measurement, and high-heat-flux cooling. AI-accelerated multiscale modeling, quantum heat-transport simulation, embedded cooling, advanced high-thermal-conductivity materials, and multiphysics co-design are expected to shift thermal management from experience-based practice toward intelligent, predictive design. The close integration of simulation and experiment, together with co-optimization of architecture, packaging, and cooling, will be essential for thermal management in next-generation high-performance electronic systems.

## Acknowledgments

The authors acknowledge support from the National Natural Science Foundation of China (Grant No. 12674045) and the Xi'an Jiaotong University Young Talent Program (Grant No. LX6J0240001). This work was carried out using computational resources provided by the National Supercomputer Center in Tianjin, including the Tianhe New Generation Supercomputer, and the Sugon Intelligent Computing Center in Xi'an.

## References

[1] Li Z Y, Luo H L, Jiang Y G, Liu H C, Xu L, Cao K Y, Wu H J, Gao P, Liu H 2024 *Appl. Therm. Eng.* **251** 123612

[2] Wang Z Q, Dong R, Ye R H, Singh S S K, Wu S F, Chen C X 2024 *Int. J. Heat Mass Transfer* **235** 126212

[3] Wei H, Ghosh S, Velusamy S, Sankaranarayanan K, Skadron K, Stan M R 2006 *IEEE Trans. VLSI Syst.* **14** 501

[4] Zhu L J, Chang K, Petranovic D, Sinha S, Yu Y S, Lim S K *2020 International Symposium on Physical Design* Taipei, China, September 20-23, 2020 p39

[5] Barber J, Sefiane K, Brutin D, Tadrist L 2009 *Appl. Therm. Eng.* **29** 1299

[6] Xia Y L, Sheng Y F, Jia R, Xu J X, Bao H 2025 *APL Electron. Device* **1** 036115

[7] Péraud J P M, Hadjiconstantinou N G 2011 *Phys. Rev. B* **84** 205331

[8] Lindsay L 2016 *Nanoscale Microscale Thermophys. Eng.* **20** 67

[9] Xiao H P, Cao W, Ouyang T, Guo S M, He C Y, Zhong J X 2017 *Sci. Rep.* **7** 45986

[10] Wang Z L, Chen G F, Zhang X L, Tang D W 2021 *Phys. Chem. Chem. Phys.* **23** 1627

[11] Yang L, Yang B S, Li B W 2023 *Phys. Rev. B* **108** 165303

[12] Wei B, Luo W, Du J, Ding Y, Guo Y, Zhu G, Zhu Y, Li B 2024 *SusMat* **4** e239

[13] Ma Z H, Zhao D, She C, Yang Y, Yang R 2021 *Mater. Today Phys.* **20** 100465

[14] Zhao D L, Qian X, Gu X K, Jajja S A, Yang R G 2016 *J. Electron. Packag.* **138** 040802

[15] Gu X K, Yang R 2014 *Appl. Phys. Lett.* **105** 131903

[16] Gu X K, Wei Y J, Yin X B, Li B W, Yang R G 2018 *Rev. Mod. Phys.* **90** 041002

[17] Liu T H, Zhou J W, Xu Q, Qian X, Song B, Yang R G 2022 *Mater. Today Phys.* **22** 100598
[18] Hu S, Zhao C Y, Gu X K 2024 *Int. J. Therm. Sci.* **196** 108725
[19] Yan S S, Wang Y, Tao F, Ren J 2022 *J. Phys. Chem. A* **126** 8771
[20] Li W, Carrete J, A. Katcho N, Mingo N 2014 *Comput. Phys. Commun.* **185** 1747
[21] Togo A 2023 *J. Phys. Soc. Jpn.* **92** 1
[22] Togo A, Chaput L, Tadano T, Tanaka I 2023 *J. Phys. : Condens. Matter* **35** 353001
[23] Tadano T, Gohda Y, Tsuneyuki S 2014 *J. Phys. : Condens. Matter* **26** 225402
[24] Kresse G, Furthmüller J 1996 *Phys. Rev. B* **54** 11169
[25] Kim J, Liu Y H, Luo T F, Tian Z T 2025 *ASME J. Heat Mass Transfer* **147** 030801
[26] Chu Y C, Shi J J, Miao K, Zhong Y, Sarangapani P, Fisher T S, Klimeck G, Ruan X L, Kubis T 2019 *Appl. Phys. Lett.* **115** 231601
[27] Feng T L, Zhong Y, Shi J J, Ruan X L 2019 *Phys. Rev. B* **99** 045301
[28] Lindsay L, Hua C, Ruan X L, Lee S 2018 *Mater. Today Phys.* **7** 106
[29] Polanco C A, Lindsay L 2019 *Phys. Rev. B* **99** 075202
[30] Dong H K, Fan Z Y, Qian P, Su Y J 2022 *Physica E* **144** 115410
[31] Córcoles A D, Chow J M, Gambetta J M, Rigetti C, Rozen J R, Keefe G A, Beth Rothwell M, Ketchen M B, Steffen M 2011 *Appl. Phys. Lett.* **99** 181906
[32] Chen L, Kumari N, Chen S T, Hou Y 2017 *RSC Adv.* **7** 26194
[33] Bao W, Wang Z, Tang D 2022 *Int. J. Heat Mass Transfer* **183** 122090
[34] Soleimani A, Araghi H, Zabihi Z, Alibakhshi A 2018 *Comput. Mater. Sci.* **142** 346
[35] Zeng J Z, Zhang D, Lu D H, Mo P H, Li Z Y, Chen Y X, Rynik M, Huang L A, Li Z Y, Shi S C, Wang Y Z, Ye H T, Tuo P, Yang J B, Ding Y, Li Y F, Tisi D, Zeng Q Y, Bao H, Xia Y, Huang J M, Muraoka K, Wang Y B, Chang J H, Yuan F B, Bore S L, Cai C, Lin Y N, Wang B, Xu J Y, Zhu J X, Luo C X, Zhang Y Z, Goodall R E A, Liang W S, Singh A K, Yao S K, Zhang J C, Wentzcovitch R, Han J Q, Liu J, Jia W L, York D M, E W, Car R, Zhang L F, Wang H 2023 *J. Chem. Phys.* **159** 054801
[36] Zeng J Z, Zhang D, Peng A Y, et al. 2025 *J. Chem. Theory Comput.* **21** 4375
[37] Wang H, Zhang L F, Han J Q, E W 2018 *Comput. Phys. Commun.* **228** 178
[38] Fan Z Y, Wang Y Z, Ying P H, Song K K, Wang J J, Wang Y, Zeng Z Z, Xu K, Lindgren E, Rahm J M, Gabourie A J, Liu J H, Dong H K, Wu J Y, Chen Y, Zhong Z, Sun J, Erhart P, Su Y J, Ala-Nissila T 2022 *J. Chem. Phys.* **157** 114801

[39] Zhou X Y, Liu Y Q, Tang B R, Wang J Y, Dong H K, Xiu X M, Chen S D, Fan Z Y 2025 *J. Appl. Phys.* **137** 014305

[40] Dong H K, Shi Y B, Ying P H, Xu K, Liang T, Wang Y Z, Zeng Z Z, Wu X, Zhou W J, Xiong S Y, Chen S D, Fan Z Y 2024 *J. Appl. Phys.* **135** 161101

[41] Li C Y, Zhang T C, Bao H G, Cheng A Q, Chen D J, Wang C F, Ding D Z, Werner D H 2025 *IEEE Trans. Microwave Theory Tech.* **73** 812

[42] Li Q S, Liu F, Hu S, Song H F, Yang S S, Jiang H L, Wang T, Koh Y K, Zhao C Y, Kang F Y, Wu J Q, Gu X K, Sun B, Wang X Q 2022 *Nat. Commun.* **13** 4901

[43] Zhou H, Zhu H L, Cui T, Pan D Z, Zhou D, Zeng X 2018 *IEEE Trans. VLSI Syst.* **26** 1312

[44] Gabourie A J, Polanco C A, McClellan C J, Su H, Malakoutian M, Çağil K, Chowdhury S, Donadio D, Pop E 2024 *IEEE International Electron Devices Meeting* San Francisco, USA, December 7-11, 2024 p1

[45] Peng J Z, Aubry N, Li Y B, Mei M, Chen Z H, Wu W T 2023 *Int. J. Heat Mass Transfer* **216** 124593

[46] Zhao X Y, Gong Z Q, Zhang Y Y, Yao W, Chen X Q 2023 *Eng. Appl. Artif. Intell.* **117** 105516

[47] Manavi S, Becker T, Fattahi E 2023 *Int. Commun. Heat Mass Transfer* **142** 106662

[48] Cai S Z, Wang Z C, Wang S F, Perdikaris P, Karniadakis G E 2021 *J. Heat Transfer* **143** 125089

[49] Sanchis-Alepuz H, Stipsitz M 2022 *IEEE Design Methodologies Conference* Bath, United Kingdom, September 1-2, 2022 p1

[50] Lu X Y, Wang Y 2024 *Proc. Combust. Inst.* **40** 105282

[51] Meethal R E, Kodakkal A, Khalil M, Ghantasala A, Obst B, Bletzinger K U, Wüchner R 2023 *Adv. Model. Simul. Eng. Sci.* **10** 6

[52] Mejail M 2025 *Int. Acad. J. Innov. Res.* **12** 43

[53] Reiser P, Aguilar J E, Guthke A, Bürkner P C 2025 *Stat. Comput.* **35** 66

[54] Westermann P, Evins R 2021 *Energy AI* **3** 100039

[55] Nemani V, Biggio L, Huan X, Hu Z, Fink O, Tran A, Wang Y, Zhang X, Hu C 2023 *Mech. Syst. Signal Process.* **205** 110796

[56] Li Y F, Xiang Y Y, Shi L J, Pan B S 2024 *J. Zhejiang Univ. Sci. A* **25** 922

[57] Fedorets A A, Dombrovsky L A, Smirnov A M 2015 *Infrared Phys. Technol.* **69** 238

[58] Cahill D G, Goodson K, Majumdar A 2001 *J. Heat Transfer* **124** 223

[59] Thorne S A, Ippolito S B, Ünlü M S, Goldberg B B 2002 *MRS Online Proc. Libr.* **738** 129

[60] Kim S H, Noh J, Jeon M K, Kim K W, Lee L P, Woo S I 2006 *J. Micromech. Microeng.* **16** 526
[61] Zhang Y C, Ding C X, Feng R, Bi K X, Geng W P, Chou X J 2024 *Measurement* **226** 114125
[62] Li X, Li Z, Zhou W, Duan Z M 2020 *IEEE Trans. VLSI Syst.* **28** 2328
[63] Cahill D G 1990 *Rev. Sci. Instrum.* **61** 802
[64] Dames C 2013 *Annu. Rev. Heat Transfer* **16** 7
[65] Zhao Y S, Liu D, Chen J, et al. 2017 *Nat. Commun.* **8** 15919
[66] Cahill D G 2018 *MRS Bull.* **43** 782
[67] Jiang P Q, Qian X, Yang R G 2018 *J. Appl. Phys.* **124** 161103
[68] Feser J P, Cahill D G 2012 *Rev. Sci. Instrum.* **83** 104901
[69] Wilson R B, Apgar B A, Martin L W, Cahill D G 2012 *Opt. Express* **20** 28829
[70] Jiang P Q, Huang B, Koh Y K 2016 *Rev. Sci. Instrum.* **87** 075101
[71] Atulasimha J, Akhras G, Flatau A B 2008 *J. Appl. Phys.* **103** 07B336
[72] Schmidt A J, Cheaito R, Chiesa M 2009 *Rev. Sci. Instrum.* **80** 094901
[73] Regner K T, Sellan D P, Su Z, Amon C H, McGaughey A J H, Malen J A 2013 *Nat. Commun.* **4** 1640
[74] Yang J, Maragliano C, Schmidt A J 2013 *Rev. Sci. Instrum.* **84** 104904
[75] Olson D H, Braun J L, Hopkins P E 2019 *J. Appl. Phys.* **126** 150901
[76] Mao R L, He P Y, Liu F C, Shi R C, Du J L, Gao P 2025 *ACS Nano* **19** 20269
[77] Xu Q, Liu W Z, Li W S, Yao T, Chu X N, Guo L J 2019 *Int. J. Heat Mass Transfer* **145** 118779
[78] Wu Z H, Xiao W, He H Y, Wang W, Song B 2025 *Nat. Electron.* **8** 810
[79] Islam S, Abdel-Motaleb I *2020 IEEE International Conference on Electro Information Technology* Chicago, USA, July 31-August 1, 2020 p233
[80] Dhumal A R, Kulkarni A P, Ambhore N H 2023 *J. Eng. Appl. Sci.* **70** 140
[81] Chen G 2021 *Nat. Rev. Phys.* **3** 555
[82] Qi Z Y, Meyer B H, Huang W, Ribando R J, Skadron K, Stan M R 2010 *IEEE International Conference on Computer Design* Amsterdam, Netherlands, October 3-6, 2010 p384
[83] Ao C, Xu B, Wang X, Chen Z Q 2025 *J. Energy Storage* **135** 118312
[84] Kim M, Kim J, Park W, Kang J S 2025 *Microelectron. Reliab.* **170** 115782
[85] Pfeifer T W, Schonfeld H B, Scott E A, Aller H T, Gaskins J T, Olson D H, Braun J L, Graham S, Hopkins P E 2025 *Annu. Rev. Mater. Res.* **55** 37
[86] Zhang C B, Wang H J, Huang Y P, Zhang L L, Chen Y P 2025 *Renew. Sustain. Energy Rev.* **208** 114989

[87] Rangarajan S, Schiffres S N, Sammakia B 2023 *Engineering* **26** 185

[88] Yu X, Shiau S-E, Ai X, Zeng Z, Zhang Z 2023 *60th ACM/IEEE Design Automation Conference* （*DAC*） San Francisco, CA, USA, July 09-13, 2023 pp1-6

[89] Chen G 2005 *Nanoscale Energy Transport and Conversion: A Parallel Treatment of Electrons, Molecules, Phonons, and Photons* （Oxford: Oxford University Press） pp1-531

[90] Sarada Devi C H, Sahaaya Arul Mary S A, Karthikeyan N, Varalakshmi S, Talasila V, Rama Naidu G 2024 *Therm. Sci. Eng. Prog.* **55** 102983

[91] Chen B Y G 2001 *Microscale Thermophys. Eng.* **5** 107

[92] Yu X Y, Chen G, Verma A, Smith J S 1995 *Appl. Phys. Lett.* **67** 3554

[93] Yao T 1987 *Appl. Phys. Lett.* **51** 1798

[94] Capinski W S, Maris H J, Ruf T, Cardona M, Ploog K, Katzer D S 1999 *Phys. Rev. B* **59** 8105

[95] Zhou J H, Zhou F, Zhao Q, Lu M X, Li Q, Chen X M 2025 *Int. Commun. Heat Mass Transfer* **161** 108555

[96] Cai Q R, Scullion D, Gan W, Falin A, Zhang S Y, Watanabe K, Taniguchi T, Chen Y, Santos E J G, Li L H 2019 *Sci. Adv.* **5** eaav0129

[97] Kim J J, Brown A D, Bakis C E, Smith E C 2021 *Compos. Sci. Technol.* **207** 108712

[98] Sun H B, Jiang Y L, Hua R J, Huang R H, Shi L, Dong Y, Liang S X, Ni J, Zhang C, Dong R Y, Song Y R 2022 *Nanomaterials* **12** 4057

[99] Baghini M S, Parvizi R, Walton F, Heidari H 2024 *IEEE Trans. Electron Devices* **71** 4203

[100] Ding T, Chen X X, Li Z Y, Liu H C, Zhu C Y, Zhao T, Li Z, Zhang Y N, Yang J A, Zhang H N, Hou L Y 2025 *Renew. Sustain. Energy Rev.* **215** 115615

[101] Al-Nagdy A A, Khalaf-Allah R A, Mohamed S M, Saeed E, Abdelaziz G B 2025 *J. Therm. Anal. Calorim.* **150** 20481

[102] Fan S M, Duan F 2020 *Int. J. Heat Mass Transfer* **150** 119324

[103] Deng D X, Zeng L, Sun W 2021 *Int. J. Heat Mass Transfer* **175** 121332

[104] Li P K, Zou Q F, Liu X L, Yang R G 2024 *Natl. Sci. Rev.* **11** nwae090

[105] Chen Y N, Zhao D Y, Liu F, Gao J, Zhu H 2023 *Microelectron. J.* **139** 105882

[106] Salvi S S, Jain A 2021 *IEEE Trans. Compon. Packag. Manuf. Technol.* **11** 802

[107] Guo Y L, Wang S 2025 *Int. Commun. Heat Mass Transfer* **162** 108592

[108] Chen H X, Sun Y, Li L H, Wang X D 2020 *Int. J. Heat Mass Transfer* **163**

120502
[109] Chakraborty B, Gallo M, Marengo M, De Coninck J, Casciola C M, Miche N, Georgoulas A 2024 *Int. J. Thermofluids* **22** 100683
[110] Hu Y, Gao H T, Yan Y Y 2025 *J. Ind. Eng. Chem.* **143** 123
[111] Kuberan V, Gedupudi S 2025 *Int. J. Heat Mass Transfer* **249** 127163
[112] Wang J J, Kaneko A, Shen B 2026 *Appl. Therm. Eng.* **283** 129000